\documentclass[aps,prb,10pt,twocolumn,floatfix,superscriptaddress,nofootinbib,longbibliography,notitlepage]{revtex4-2}

\usepackage{iftex}
\ifPDFTeX
  \usepackage[T1]{fontenc}
\fi
\usepackage{amsmath,amssymb}
\usepackage{booktabs}
\usepackage{bm}
\usepackage{graphicx}
\usepackage[colorlinks=true,citecolor=blue,linkcolor=red]{hyperref}
\usepackage{listings}
\usepackage{xcolor}
\usepackage{tikz}
\usetikzlibrary{arrows.meta,positioning}
\hypersetup{breaklinks=true}
\makeatletter
\renewcommand{\frontmatter@thefootnote}{  \@fnsymbol{\numexpr\csname c@\@mpfn\endcsname+1\relax}}
\makeatother
\newcommand{\moirekp}{\texttt{MoireKP}}
\newcommand{\heff}{H_{\mathrm{eff}}}

\newcommand{\Ht}{H^{\mathrm{TAPW}}}
\newcommand{\St}{S^{\mathrm{TAPW}}}
\newcommand{\Horth}{H_{\mathrm{TAPW}}^{\mathrm{orth}}}
\newcommand{\Hcont}{H_{\mathrm{cont}}}
\newcommand{\kvec}{\mathbf{k}}
\newcommand{\bQ}{\mathbf{Q}}
\newcommand{\bG}{\mathbf{G}}

\newcommand{\bR}{\mathbf{R}}

\newcommand{\PaperStat}[2]{\csname paperstat#1#2\endcsname}

\expandafter\def\csname paperstatmote2_kraw\endcsname{0.7}
\expandafter\def\csname paperstatmote2_kaligned\endcsname{0.6}
\expandafter\def\csname paperstatmote2_kmax\endcsname{2.0}
\expandafter\def\csname paperstatmote2_kshift\endcsname{0.3}
\expandafter\def\csname paperstatmote2_koverlapmean\endcsname{1.0}
\expandafter\def\csname paperstatmote2_koverlapmin\endcsname{0.8}

\expandafter\def\csname paperstatzrs2_mraw\endcsname{1.4}
\expandafter\def\csname paperstatzrs2_maligned\endcsname{0.6}
\expandafter\def\csname paperstatzrs2_mmax\endcsname{2.5}
\expandafter\def\csname paperstatzrs2_mshift\endcsname{-1.3}
\expandafter\def\csname paperstatzrs2_moverlapmean\endcsname{0.9}
\expandafter\def\csname paperstatzrs2_moverlapmin\endcsname{0.9}

\expandafter\def\csname paperstatzrs2_gammaraw\endcsname{0.6}
\expandafter\def\csname paperstatzrs2_gammaaligned\endcsname{0.6}
\expandafter\def\csname paperstatzrs2_gammamax\endcsname{2.7}
\expandafter\def\csname paperstatzrs2_gammashift\endcsname{-0.2}
\expandafter\def\csname paperstatzrs2_gammaoverlapmean\endcsname{1.0}
\expandafter\def\csname paperstatzrs2_gammaoverlapmin\endcsname{1.0}

\expandafter\def\csname paperstataab_karaw\endcsname{2.5}
\expandafter\def\csname paperstataab_kaaligned\endcsname{1.0}
\expandafter\def\csname paperstataab_kamax\endcsname{5.3}
\expandafter\def\csname paperstataab_kashift\endcsname{2.3}
\expandafter\def\csname paperstataab_kaoverlapmean\endcsname{0.9}
\expandafter\def\csname paperstataab_kaoverlapmin\endcsname{0.5}

\expandafter\def\csname paperstataab_kbraw\endcsname{3.1}
\expandafter\def\csname paperstataab_kbaligned\endcsname{1.1}
\expandafter\def\csname paperstataab_kbmax\endcsname{4.8}
\expandafter\def\csname paperstataab_kbshift\endcsname{2.3}
\expandafter\def\csname paperstataab_kboverlapmean\endcsname{0.8}
\expandafter\def\csname paperstataab_kboverlapmin\endcsname{0.5}

\expandafter\def\csname paperstataab_gammaraw\endcsname{0.6}
\expandafter\def\csname paperstataab_gammaaligned\endcsname{0.5}
\expandafter\def\csname paperstataab_gammamax\endcsname{2.6}
\expandafter\def\csname paperstataab_gammashift\endcsname{0.4}
\expandafter\def\csname paperstataab_gammaoverlapmean\endcsname{1.0}
\expandafter\def\csname paperstataab_gammaoverlapmin\endcsname{0.9}
\expandafter\def\csname paperstatmote2_kdisplayrms\endcsname{0.8}
\expandafter\def\csname paperstatmote2_kdisplaymax\endcsname{2.0}
\expandafter\def\csname paperstatzrs2_mdisplayrms\endcsname{0.6}
\expandafter\def\csname paperstatzrs2_mdisplaymax\endcsname{1.4}
\expandafter\def\csname paperstatzrs2_gammadisplayrms\endcsname{0.6}
\expandafter\def\csname paperstatzrs2_gammadisplaymax\endcsname{2.7}
\expandafter\def\csname paperstataab_kadisplayrms\endcsname{1.0}
\expandafter\def\csname paperstataab_kadisplaymax\endcsname{3.0}
\expandafter\def\csname paperstataab_kbdisplayrms\endcsname{1.1}
\expandafter\def\csname paperstataab_kbdisplaymax\endcsname{2.5}
\expandafter\def\csname paperstataab_gammadisplayrms\endcsname{0.6}
\expandafter\def\csname paperstataab_gammadisplaymax\endcsname{2.6}

\begin{document}
\sloppy

\title{MoireKP: An Automated Workflow for Symmetry-Constrained Moiré Continuum Models}

\author{Yan Zhang}
\affiliation{Beijing National Laboratory for Condensed Matter Physics and Institute of Physics,
Chinese Academy of Sciences, Beijing 100190, China}
\affiliation{University of Chinese Academy of Sciences, Beijing 100049, China}

\author{Hanqi Pi}
\affiliation{Donostia International Physics Center (DIPC), Paseo Manuel de Lardizábal 4,
20018 Donostia-San Sebastián, Spain}

\author{Wangqian Miao}
\affiliation{Department of Physics, The Hong Kong University of Science and Technology,
Clear Water Bay, Hong Kong, China}

\author{Zhong Fang}
\affiliation{Beijing National Laboratory for Condensed Matter Physics and Institute of Physics,
Chinese Academy of Sciences, Beijing 100190, China}
\affiliation{University of Chinese Academy of Sciences, Beijing 100049, China}

\author{Hongming Weng}
\affiliation{Beijing National Laboratory for Condensed Matter Physics and Institute of Physics,
Chinese Academy of Sciences, Beijing 100190, China}
\affiliation{University of Chinese Academy of Sciences, Beijing 100049, China}
\affiliation{Condensed Matter Physics Data Center of Chinese Academy of Sciences,
Beijing 100190, China}

\author{Xi Dai}
\affiliation{Department of Physics, The Hong Kong University of Science and Technology,
Clear Water Bay, Hong Kong, China}

\author{B. Andrei Bernevig}
\affiliation{Department of Physics, Princeton University, Princeton, New Jersey 08544, USA}
\affiliation{Donostia International Physics Center (DIPC), Paseo Manuel de Lardizábal 4,
20018 Donostia-San Sebastián, Spain}
\affiliation{IKERBASQUE, Basque Foundation for Science, Bilbao, Spain}

\author{Quansheng Wu}
\email{quansheng.wu@iphy.ac.cn}
\affiliation{Beijing National Laboratory for Condensed Matter Physics and Institute of Physics,
Chinese Academy of Sciences, Beijing 100190, China}
\affiliation{University of Chinese Academy of Sciences, Beijing 100049, China}

\date{October 28, 2026}

\begin{abstract}
First-principles Hamiltonians contain the electronic information needed to study moir\'e materials, but their size makes subsequent calculations costly. We present \moirekp{}, an automated workflow for constructing compact continuum models from localized-orbital \textit{ab initio} Hamiltonians. The workflow constructs a valley-resolved truncated atomic-plane-wave (TAPW) Hamiltonian, selects a low-energy basis, and projects the crystal-symmetry operations into this basis. These representations determine the symmetry-allowed Hamiltonian terms, whose coefficients are obtained by projecting the low-energy effective Hamiltonian onto the corresponding operator basis.
The model dimension, localized reference states, and harmonic cutoff can be selected automatically or specified by the user. This Hamiltonian-based construction retains wave-function information within the selected low-energy subspace. We apply the workflow to AA-stacked twisted bilayer MoTe$_2$, AB-stacked twisted bilayer ZrS$_2$, and A--AB twisted trilayer MoTe$_2$, covering K, M, and \(\Gamma\) valleys. All models reproduce TAPW band energies within 3.0 meV in the target windows along high-symmetry \(k\) paths.

\end{abstract}

\maketitle

\section{Introduction}

A small twist or lattice mismatch between van der Waals layers creates a long-period moiré superlattice whose modulation can produce narrow bands with tunable bandwidth, topology, and quantum geometry~\cite{bistritzer_Moire_2011,andrei_Marvels_2021,carr_Electronicstructure_2020,wu_topological_2019,devakul_magic_2021}. Modelling systems with different valley geometries, orbital manifolds, and layer configurations calls for a common procedure for constructing low-energy bases and their symmetry representations. This need extends beyond the familiar \(K\)-valley setting to \(M\)-valley~\cite{calugaru_Moire_2025,lei_Moire_2025}, \(\Gamma\)-valley~\cite{angeli_Valley_2021,foutty_Tunable_2023,pi_engineering_2026,li_Geometric_2026}, and multivalley and multilayer systems~\cite{qi_ChernSelective_2025,kuklin_Contrasting_2026}. Band energies alone do not determine the wave functions, which govern band topology and quantum geometry~\cite{zhang_universal_2024,devakul_magic_2021,wang_fractional_2024} and enter the form factors of projected interactions~\cite{reddy_fractional_2023,yu_fractional_2024}. Real-space measurements of layer polarization in twisted MoTe$_2$ further connect wave-function structure to band topology~\cite{thompson_Microscopic_2025}. These considerations motivate low-energy Hamiltonians that are not tied to a prescribed material-specific form and retain the band dispersion together with wave-function and symmetry information.

\begin{figure*}[t]
  \centering
  \includegraphics[width=0.98\textwidth]{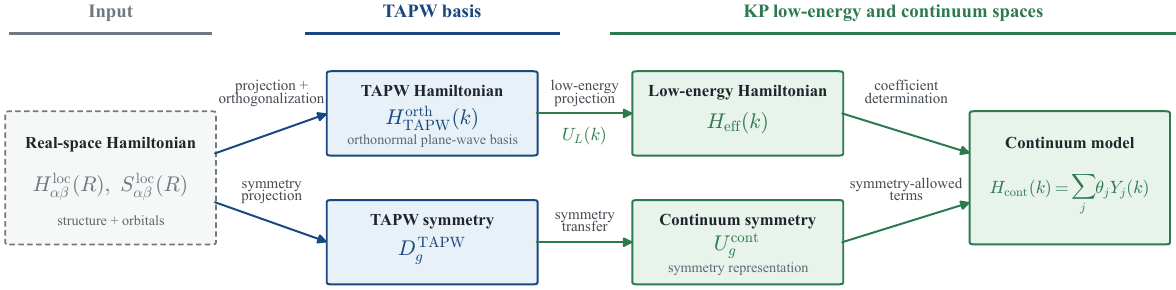}
  \caption{First-principles-to-continuum construction implemented in MoireKP. Starting from the localized-orbital Hamiltonian, overlap matrix, and structural data, the program constructs the orthonormalized TAPW Hamiltonian and its symmetry matrices, selects the low-energy basis, and obtains the coefficients of the symmetry-allowed Hamiltonian terms from \(H_{\rm eff}\).}
  \label{fig:workflow}
\end{figure*}

First-principles calculations provide the spectral, wave-function, and symmetry information, but the number of localized orbitals grows with the moiré supercell. Repeated use of the resulting Hamiltonian in topological, quantum-geometric, and many-body calculations is therefore expensive. Continuum models reduce the problem to the relevant degrees of freedom near selected valleys and provide a compact Hamiltonian for subsequent calculations~\cite{lopesdossantos_Graphene_2007,balents_General_2019,vafek_Continuum_2023}. Constructing such a model nevertheless requires identifying the relevant low-energy basis, determining the valley symmetry representation in that basis, and specifying the symmetry-allowed  Hamiltonian terms.

Complementary approaches to moiré electronic structure include Wannier-derived atomistic Hamiltonians~\cite{fang_Electronic_2016}, large-scale first-principles calculations~\cite{zhang_Initio_2022}, and machine-learned Hamiltonians~\cite{li_DeepH_2022,gong_DeepHE3_2023}. Microscopic constructions also derive continuum Hamiltonians from atomistic or first-principles descriptions~\cite{carr_Exact_2019,rost_Nonperturbative_2019,cances_Simple_2023}.
Many material-specific approaches specify the low-energy fields and continuum Hamiltonian before determining their parameters from band structures or stacking-dependent calculations~\cite{jung_Initio_2014,angeli_Valley_2021,mao_Transfer_2024}. More recent approaches have made continuum models transferable across twist angles by combining angle-dependent lattice relaxation with electronic parameters extracted from first-principles calculations~\cite{zhang_transferable_2025,xie_generic_2025}. MoireStudio integrates moiré-structure generation with tight-binding and parameterized continuum-model calculations~\cite{yu_MoireStudio_2026}.

The truncated atomic-plane-wave (TAPW) method projects the localized-orbital Hamiltonian of a large moiré supercell onto a reduced basis of atomic plane waves whose momenta lie near a selected valley~\cite{miao_truncated_2023,shi_tbg_epc_2024}. Using this representation, Ref.~\cite{zhang_universal_2024} introduced a Hamiltonian-based construction of continuum models without fitting band energies. The method selects a low-energy subspace, expresses the symmetry-allowed Hamiltonian terms as matrices
in the plane-wave basis, and obtains their coefficients by matrix-trace projection of a low-energy effective Hamiltonian. In those applications, however, the low-energy states, valley-symmetry representation matrices, and symmetry-allowed continuum Hamiltonian terms vary among systems and require manual work.

Building on the Hamiltonian-based construction of Ref.~\cite{zhang_universal_2024}, here we present \moirekp{}, which automates these model-dependent steps. The workflow selects the low-energy subspace, fixes a physically interpretable basis with localized reference states, and projects the crystal-symmetry operations into this basis. The resulting representation matrices determine the independent symmetry-allowed Hamiltonian terms. The coefficients multiplying these terms are obtained by matrix-trace projection of the low-energy effective Hamiltonian derived from the TAPW Hamiltonian.
The required inputs are the nonorthogonal Hamiltonian and overlap matrices, crystal structure, and orbital information. \moirekp{} supports these inputs from OpenMX~\cite{ozaki_Variationally_2003,ozaki_Numerical_2004,ozaki_Efficient_2005}, SIESTA~\cite{soler_SIESTA_2002}, and ABACUS~\cite{li_Largescale_2016,chen_Systematically_2010}. Relaxation and corrugation are included through first-principles calculations on relaxed structures.

We apply the same construction to the \(K\) valley of AA-stacked twisted bilayer MoTe$_2$, the \(M\) and \(\Gamma\) valleys of AB-stacked twisted bilayer ZrS$_2$, and the \(K_A\), \(K_B\), and \(\Gamma\) sectors of A--AB twisted trilayer MoTe$_2$. We benchmark the resulting models against TAPW using band dispersions, wave-function overlaps, and selected model-space geometric quantities.

\section{Results}\label{sec:results}

\begin{figure*}[!t]
  \centering
  \includegraphics[width=0.98\textwidth]{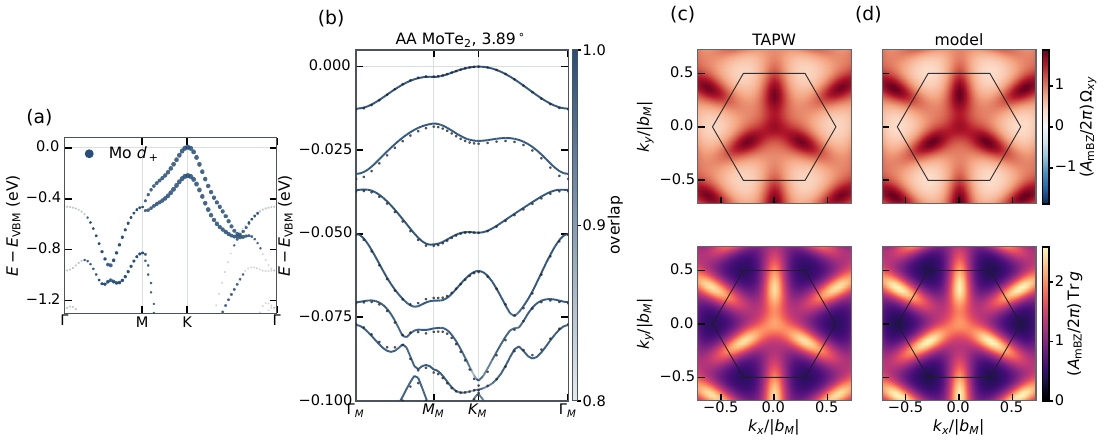}
  \caption{\(K\)-valley continuum model for AA-stacked twisted bilayer MoTe$_2$ at \(3.89^\circ\). (a) Mo \(d_+\)-projected monolayer valence bands; marker size gives orbital weight. (b) TAPW bands (black dots) and continuum bands colored by wave-function overlap [Sec.~\ref{sec:validation_protocol}]. (c,d) TAPW and continuum-model Berry curvature (top) and quantum-metric trace (bottom) for the highest valence band. Energies in (a) use the monolayer VBM; in (b), the model VBM is aligned to the TAPW VBM, which sets zero. Hexagons mark the first moir\'e Brillouin zone.
    }
  \label{fig:mote2_k_flagship}
\end{figure*}

\subsection{Overview of MoireKP}\label{sec:architecture}

Figure~\ref{fig:workflow} separates the construction into two branches that meet in the final continuum model. \moirekp{} starts from a real-space localized-orbital Hamiltonian, overlap matrix, crystal structure, and orbital basis. For the systems presented here, these inputs are obtained from OpenMX calculations. After the user specifies the target valley, TAPW projection and symmetric orthogonalization give \(\Horth\). The workflow then selects the low-energy subspace, fixes its basis \(U_L\) using localized reference states, and constructs \(H_{\rm eff}\). Its matrix elements are used to determine the continuum-model parameters.

The second branch constructs \(D_g^{\rm TAPW}\), the matrix representing the symmetry action in the TAPW basis. For an antiunitary operation, the action also complex conjugates the state coefficients. The TAPW symmetry construction uses the rigid reference structure; real-space H/S averaging uses operations identified from the relaxed input structure with \texttt{spglib}~\cite{togo_spglib_2024}. Time reversal and its combinations with spatial operations are retained only when they are symmetries of the input Hamiltonian and permute the selected valley sectors and plane-wave channels. Within each sector, the same ordered reference-state pattern is used for every selected \(\bQ\) channel. When the selected subspace and the reference states satisfy the symmetry-closure conditions in Appendix~\ref{app:low_energy_symmetry}, the projected symmetry action separates into a channel permutation and a small internal matrix \(\rho_g\) acting on the layer, orbital, and spin labels. Together, they define the matrix part \(U_g^{\rm cont}\) of the continuum symmetry representation, which constrains the allowed Hamiltonian terms. The continuum Hamiltonian is
\begin{equation}
  H_{\rm cont}(\kvec)=\sum_j\theta_jY_j(\kvec),
  \label{eq:overview_continuum_hamiltonian}
\end{equation}
where \(Y_j(\kvec)\) are the symmetry-allowed Hamiltonian terms within the chosen momentum-expansion and harmonic truncation, and \(\theta_j\) are their coefficients. The matrices \(U_g^{\rm cont}\) constrain the allowed terms, while matrix-trace projection of \(H_{\rm eff}\) determines \(\theta_j\).

\begin{figure*}[!ht]
  \centering
  \includegraphics[width=0.98\textwidth]{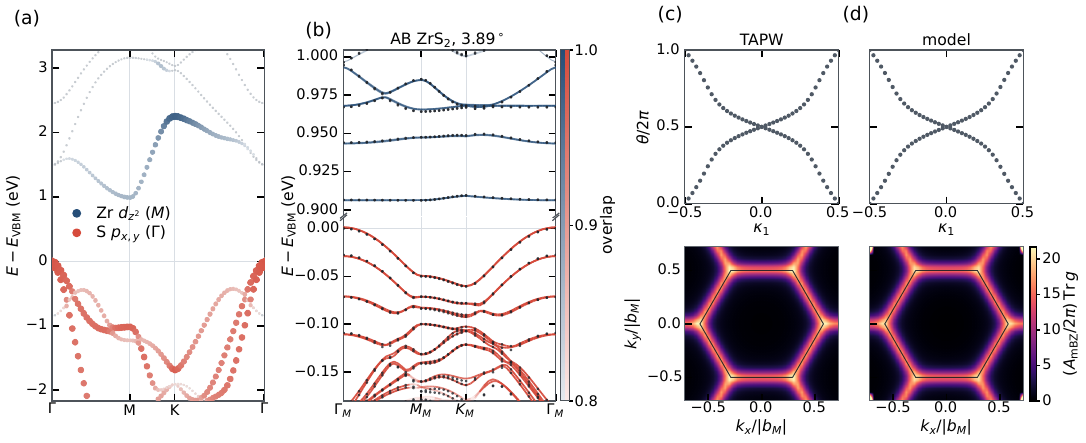}
  \caption{\(M\)- and \(\Gamma\)-valley continuum models for AB-stacked twisted ZrS$_2$ at \(3.89^\circ\). (a) Monolayer bands projected onto Zr \(d_{z^2}\) near \(M\) and S \(p_x/p_y\) near \(\Gamma\). (b) TAPW bands (black dots) and continuum bands for \(M\) (top, blue) and \(\Gamma\) (bottom, red), shaded by wave-function overlap. The $\Gamma$ error and overlap statistics use the highest ten bands. (c,d) TAPW and continuum-model Wilson-loop eigenphases along \(\mathbf b_2\) (top) and quantum-metric trace (bottom) for the highest two \(\Gamma\) valence bands. Energies in (a) use the monolayer VBM; both valleys in (b) use the TAPW \(\Gamma\) VBM, with the \(M\)-model CBM aligned to TAPW. Overlaps and geometric quantities are defined in Sec.~\ref{sec:validation_protocol}. Hexagons mark the first moir\'e Brillouin zone.
    }
  \label{fig:zrs2_multivalley}
\end{figure*}

Figures~\ref{fig:mote2_k_flagship}--\ref{fig:mote2_aab_multivalley} compare the continuum and TAPW band dispersions and quantify their wave-function overlaps. Berry curvature, quantum metric, and Wilson loops are also compared for selected bands or subspaces. Model configurations and validation results are summarized in Table~\ref{tab:paper_model_configurations}.

The examples differ in their low-energy degrees of freedom and basis selection. The bilayer \(K\) model retains one orbital state per layer. The \(M\)-valley states are selected separately in the two layers, whereas the \(\Gamma\)-valley states are selected jointly to include same-\(Q\) interlayer hybridization. The trilayer sectors retain different combinations of three physical layers.

\subsection{\texorpdfstring{\(K\)-valley model in AA-stacked twisted bilayer MoTe$_2$}{K-valley benchmark in AA-stacked twisted bilayer MoTe2}}

Twisted bilayer MoTe$_2$ is a well-studied \(K\)-valley moir\'e system. Continuum modelling and first-principles calculations have found narrow topological moir\'e valence bands~\cite{wu_topological_2019,jia_moire_2024,wang_fractional_2024,mao_Transfer_2024}, and experiments have reported integer and fractional Chern states in the same material~\cite{cai_Signatures_2023,zeng_Thermodynamic_2023,xu_Observation_2023,park_Observation_2023}. We use AA-stacked twisted bilayer MoTe$_2$ at \(3.89^\circ\) as the first benchmark. The AA-stacked twisted bilayer MoTe$_2$ has \(C_{3z}\) and an in-plane \(C_2\) as crystalline symmetries. In the single-valley model constructed here, \(C_{3z}\) and the antiunitary combination \(C_2\mathcal T\) act within the selected \(K\) valley, whereas \(C_2\) and \(\mathcal T\) exchange the \(K\) and \(K'\) valley~\cite{zhang_universal_2024}. We use \(K'\equiv-K\) modulo a monolayer reciprocal-lattice vector.

In monolayer MoTe$_2$, the valence-band maximum at the selected \(K\) valley is dominated by the Mo \(d_+=(d_{x^2-y^2}+\mathrm{i}d_{xy})/\sqrt2\) orbital~\cite{wu_topological_2019,jia_moire_2024}. The continuum basis for the twisted bilayer therefore contains the corresponding \(d_+\)-dominated state centered at the rotated \(K\) valley of each layer.

For the highest eight bands along the high-symmetry \(k\) path, the RMS energy difference from TAPW after alignment is \(\PaperStat{mote2_k}{displayrms}\) meV, with the corresponding eigenstate overlaps shown in Fig.~\ref{fig:mote2_k_flagship}(b).

\subsection{\texorpdfstring{\(M\)- and \(\Gamma\)-valley models in AB-stacked twisted ZrS$_2$}{M- and Gamma-valley models in AB-stacked twisted ZrS2}}
\label{sec:zrs2_multivalley}

We next apply \moirekp{} to the \(M\)- and \(\Gamma\)-valley states in AB-stacked twisted bilayer ZrS$_2$ which is a platform with two types of valley. In monolayer 1T-ZrS$_2$, the conduction-band minimum lies at \(M\)~\cite{calugaru_Moire_2025,lei_Moire_2025} and the valence-band maximum lies at \(\Gamma\)~\cite{claassen_UltraStrong_2022,pi_engineering_2026}. The three \(M\) valleys are related by \(C_{3z}\), and each is invariant under time reversal~\cite{calugaru_Moire_2025,lei_Moire_2025}. Here, we construct the model for \(M_1\) valley and the other two \(M\) valleys can be obtained by \(C_{3z}\). Within each selected \(M\) valley, the corresponding continuum model has the time-reversal symmetry \(\mathcal T\) and an in-plane \(C_2\), which exchanges top and bottom layer. For \(\Gamma\)-valley, the valley symmetries are \(\mathcal T\), \(C_{3z}\), and an in-plane \(C_2\).

For \(M_1\), \moirekp{} selects the conduction states separately in each layer and fixes their basis using Zr \(d_{z^2}\) orbital references. At the monolayer \(\Gamma\) point, the valence states are dominated by S \(p_x/p_y\) orbitals and form spinful Kramers pairs~\cite{claassen_UltraStrong_2022,pi_engineering_2026}. At \(\Gamma\), the two layers share the same valley center. We select the S \(p_x/p_y\)-dominated valence states jointly from a fixed-\(\bQ\) block containing both layers, so that interlayer hybridization within that block is included in the initial low-energy states.

Figure~\ref{fig:zrs2_multivalley}(b) compares the continuum and TAPW bands and their eigenstate overlaps. Along the high-symmetry \(k\) path, the RMS energy differences after alignment are \(\PaperStat{zrs2_m}{displayrms}\) meV for the lowest eight \(M_1\) bands and \(\PaperStat{zrs2_gamma}{displayrms}\) meV for the highest ten \(\Gamma\) bands.

The \(M\) and \(\Gamma\) calculations use the same construction but select the low-energy states separately and jointly, respectively.

\subsection{\texorpdfstring{Multilayer and multivalley models in A--AB trilayer MoTe$_2$}{Multilayer and multivalley models in A--AB trilayer MoTe2}}
\label{sec:aab_mote2_multivalley}

We next consider A--AB twisted trilayer MoTe$_2$, in which \(K\)- and \(\Gamma\)-valley states occur in the low-energy region. The letters A and B label layer orientations: in the untwisted AAB reference structure, the two A layers have the same orientation, while the B layer is rotated by \(180^\circ\) relative to them. The \(A_1\) monolayer is twisted relative to the \(A_2B\) bilayer, and the subscripts distinguish the two A layers~\cite{qi_ChernSelective_2025}.
Interlayer hybridization shifts the highest \(\Gamma\)-valley valence states upward compared with the twisted-bilayer case, bringing both \(\Gamma\) and \(K\) valleys into the low-energy region~\cite{qi_ChernSelective_2025}.

At \(K\), spin--valley locking gives the low-energy A- and B-layer valence states opposite dominant spin polarizations, suppressing their interlayer hybridization. Following Ref.~\cite{qi_ChernSelective_2025}, we construct the \(K_A\) and \(K_B\) models from the spin-diagonal TAPW blocks, retaining their spin--orbit coupling terms and neglecting spin-flip mixing. Relative to full-spin TAPW, this approximation changes the energies of the highest six \(K_A\) and three \(K_B\) bands by at most \(2.8\) and \(2.9\) meV, respectively, along the high-symmetry \(k\) path, after aligning each spectrum at its \(K_A\) valence-band maximum. The \(K_A\) sector contains the two A layers, and \(K_B\) contains the B layer. These states are dominated by the Mo \(d_{\pm2}\equiv d_\pm=(d_{x^2-y^2}\pm\mathrm{i}d_{xy})/\sqrt2\) orbitals. The \(\Gamma\)-valley model instead uses Mo \(d_{z^2}\)-dominated states.

The A--AB structure breaks the in-plane \(C_2\) symmetry of the twisted bilayer. Both \(K_A\) and \(K_B\) models retain \(C_{3z}\), while time reversal connects each model to its own opposite-valley partner. At \(\Gamma\), both time reversal and \(C_{3z}\) act within the model~\cite{qi_ChernSelective_2025}.

\begin{figure*}[!t]
  \centering
  \includegraphics[width=0.96\textwidth]{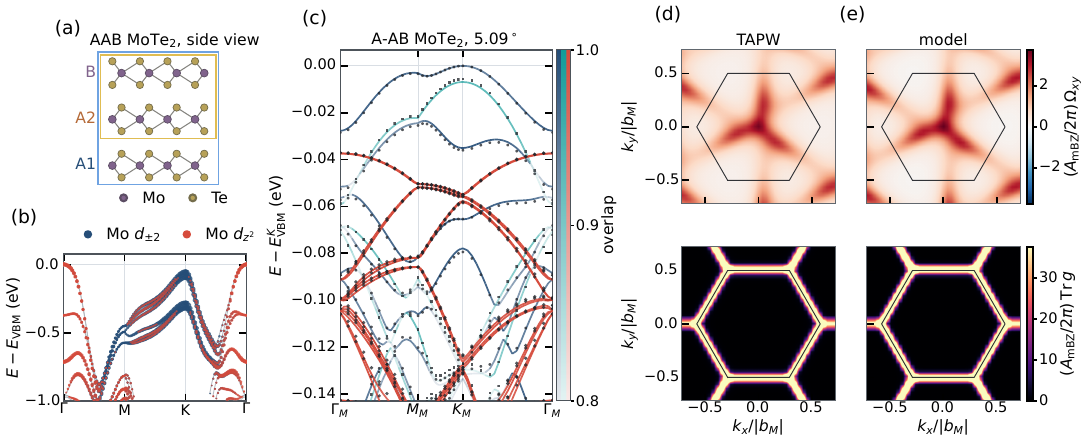}
  \caption{Continuum models for A--AB trilayer MoTe$_2$ at \(5.09^\circ\). (a) Side view of the untwisted \(A_1\)--\(A_2\)--\(B\) structure. (b) Valence bands of the relaxed untwisted trilayer; marker size gives Mo \(d_{\pm2}\) and \(d_{z^2}\) weights. (c) TAPW bands (black symbols) and continuum bands colored by wave-function overlap [Sec.~\ref{sec:validation_protocol}]. (d,e) TAPW and continuum-model Berry curvature of the highest \(K_A\) band (top) and quantum-metric trace of the two-band \(\Gamma\) subspace (bottom). Energies in (b) use the untwisted-trilayer VBM. In (c), zero is the TAPW \(K_A\) VBM; the \(K_A/K_B\) model bands share the offset aligning the \(K_A\) VBM to TAPW. Hexagons mark the first moir\'e Brillouin zone.
    }
  \label{fig:mote2_aab_multivalley}
\end{figure*}

The orbital weights in Fig.~\ref{fig:mote2_aab_multivalley}(b) distinguish the Mo \(d_{\pm2}\)-dominated \(K\)-valley states from the Mo \(d_{z^2}\)-dominated \(\Gamma\)-valley states. For the \(K_A\) block, \moirekp{} selects Mo \(d_{\pm2}\) states from the \(A_1\) and \(A_2\) layers; for the \(K_B\) block, it selects the corresponding state from the B layer. At \(\Gamma\), four low-energy states are selected jointly from the same-\(Q\) block containing all three layers, with Mo \(d_{z^2}\) orbitals on \(A_1\) and \(A_2\) serving as basis references.

The \(K_A\), \(K_B\), and \(\Gamma\) continuum models reproduce the TAPW dispersions in Fig.~\ref{fig:mote2_aab_multivalley}(c), with RMS energy differences after alignment of \(\PaperStat{aab_ka}{displayrms}\), \(\PaperStat{aab_kb}{displayrms}\), and \(\PaperStat{aab_gamma}{displayrms}\) meV for the highest six, three, and eight bands, respectively, along the high-symmetry \(k\) path.

\section{Discussion}
\label{sec:discussion}

\moirekp{} formulates continuum-model construction as a symmetry-constrained projection of a localized-orbital \textit{ab initio} Hamiltonian. Building on the Hamiltonian-projection method of Ref.~\cite{zhang_universal_2024}, \moirekp{} automates low-energy subspace and reference-state selection, projects the input symmetry representations into the selected basis, and generates the symmetry-allowed operator basis for coefficient extraction.

The target bands and spin/valley sectors remain user choices. Accuracy depends on the TAPW cutoff, selected low-energy space, polynomial and harmonic truncations, reference energy and separation from excluded states, and coefficient-extraction sampling. Reduction and expansion errors are distinct from approximations in the first-principles/TAPW input. The resulting explicit \(H_{\rm cont}\) can be evaluated independently in subsequent band, symmetry, topological, quantum-geometric, and many-body calculations.

\section{Methods}\label{sec:formulation}

The construction has two coupled parts. The first symmetrizes the localized-orbital Hamiltonian and overlap matrices, projects them to the TAPW basis, transforms the Hamiltonian to a L\"owdin-orthogonalized basis, and reduces it to an effective low-energy Hamiltonian \(H_{\rm eff}\). The second starts from the unitary and antiunitary symmetries of the localized-orbital Hamiltonian and retains only the operations that permute the selected valley sectors and plane-wave channels. Their representation matrices are projected first into the TAPW basis and then into the same low-energy basis used for \(H_{\rm eff}\). These matrices determine the symmetry-allowed Hamiltonian terms, whose coefficients are obtained by trace projection of \(H_{\rm eff}\).

The TAPW projection follows Ref.~\cite{miao_truncated_2023}. The low-energy reduction and the plane-wave representation of continuum Hamiltonian terms follow Ref.~\cite{zhang_universal_2024}.

\subsection{Localized-orbital matrices and real-space symmetry}

The OpenMX input consists of the real-space Hamiltonian and overlap matrices of the relaxed moiré structure~\cite{ozaki_Variationally_2003,ozaki_Numerical_2004,ozaki_Efficient_2005}. The calculations use fully relativistic PBE19 norm-conserving pseudopotentials and optimized pseudoatomic-orbital (PAO) basis functions from the OpenMX 2019 database~\cite{perdew_Generalized_1996,openmx_Database_2019}. The two MoTe$_2$ examples use Mo7.0-\(s3p2d2\) and Te7.0-\(s3p2d2f1\), giving 71 spatial PAOs per monolayer unit cell. The ZrS$_2$ example uses Zr7.0-\(s3p2d1\) and S7.0-\(s2p2d1\), giving 40 spatial PAOs per monolayer unit cell. In these labels, 7.0 is the PAO cutoff radius in Bohr. Let \(|\bR,p\rangle\) denote the localized-orbital basis state associated with orbital \(p\) in the moiré cell translated by \(\bR\). Their matrix elements are
\begin{equation}
  \begin{aligned}
    \relax[H^{\rm OMX}(\bR)]_{pq}
    &=
    \langle \mathbf 0,p|
    \widehat H
    |\bR,q\rangle,\\
    [S^{\rm OMX}(\bR)]_{pq}
    &=
    \langle \mathbf 0,p|\bR,q\rangle .
  \end{aligned}
  \label{eq:localized_real_space_matrices}
\end{equation}
The localized orbitals are not assumed to be orthogonal, so the overlap matrix is retained throughout the TAPW projection.

\moirekp{} also uses the corresponding rigid commensurate structure, with the same atom and orbital ordering, to assign each localized orbital to a physical layer, a monolayer unit cell, a basis atom, and an orbital--spin state. We write
\[
  \alpha=(\mu,\zeta),\qquad
  p=(\ell,i,\alpha),\qquad
  \bm\tau_p\equiv\bm\tau_{\ell i\mu}
  =\bm\rho_{\ell i}+\mathbf d_{\ell\mu},
\]
where \(\ell\) labels the physical layer, \(i\) labels a monolayer unit cell within the rigid reference moiré cell, \(\mu\) labels a basis atom, and \(\zeta\) labels its localized orbital and spin state. The vector \(\bm\rho_{\ell i}\) is the translation of monolayer cell \(i\), and \(\mathbf d_{\ell\mu}\) is the position of atom \(\mu\) within that cell. Thus, \(\bm\tau_{\ell i\mu}\) denotes this rigid reference position. It fixes the Bloch and TAPW phase convention, while relaxation and corrugation remain encoded in \(H^{\rm OMX}(\bR)\) and \(S^{\rm OMX}(\bR)\), and hence in the symmetrized matrices below.

We use \texttt{spglib}~\cite{togo_spglib_2024} to identify spatial operations of the relaxed input structure at the specified tolerance. For each admitted operation \(g\), the atom map \(u\mapsto gu\) and orbital--spin rotation \(U_{g,u}\) define the real-space representation \(\widehat g\). The present symmetry representation is restricted to Seitz operations with zero attached translation in the chosen origin convention; nonsymmorphic operations requiring irreducible fractional translations are outside the current implementation.

Before the Bloch transform, the real-space Hamiltonian and overlap matrices are symmetrized under these spatial operations. The matrices used below are

\begin{equation}
  \begin{aligned}
    \relax[H^{\rm loc}(\bR)]_{pq}
    &=\frac{1}{N_{\rm sym}}\sum_g
    [\widehat g\,H^{\rm OMX}\widehat g^\dagger]_{pq}(\bR),\\
    [S^{\rm loc}(\bR)]_{pq}
    &=\frac{1}{N_{\rm sym}}\sum_g
    [\widehat g\,S^{\rm OMX}\widehat g^\dagger]_{pq}(\bR),
  \end{aligned}
  \label{eq:real_space_symmetrization}
\end{equation}
where the sum runs over the \(N_{\rm sym}\) operations admitted by the relaxed-structure check. The matrix elements come from the relaxed OpenMX calculation. Across the four edge-band windows specified in Appendix~\ref{app:subsec:real_space_averaging_checks}, the maximum energy changes under symmetry averaging range from 0.2 to 4.6~meV along the high-symmetry \(k\) path. The explicit transformation of the real-space blocks is given in Appendix~\ref{app:subsec:local_orbital_symmetry}. The subsequent construction uses the symmetry-averaged matrices.

\subsection{TAPW projection and symmetry}

Let \(\bR_I\) be a moiré-superlattice translation and \(N_{\rm m}\) the number of moiré cells. In the orbital-centered phase convention, the localized-orbital Bloch basis is
\begin{equation}
  |\chi_p(\kvec)\rangle
  =
  \frac{1}{\sqrt{N_{\rm m}}}
  \sum_I
  e^{i\kvec\cdot(\bR_I+\bm\tau_p)}
  |\bR_I,p\rangle .
  \label{eq:localized_bloch_basis_main}
\end{equation}
The corresponding Bloch Hamiltonian and overlap matrices are
\begin{equation}
  \begin{aligned}
    \relax[H^{\rm loc}(\kvec)]_{pq}
    &=
    \sum_{\bR}
    e^{i\kvec\cdot(\bR+\bm\tau_q-\bm\tau_p)}
    [H^{\rm loc}(\bR)]_{pq},\\
    [S^{\rm loc}(\kvec)]_{pq}
    &=
    \sum_{\bR}
    e^{i\kvec\cdot(\bR+\bm\tau_q-\bm\tau_p)}
    [S^{\rm loc}(\bR)]_{pq}.
  \end{aligned}
  \label{eq:localized_bloch_matrices}
\end{equation}

\begin{figure*}[t]
  \centering
  \includegraphics[width=0.98\textwidth]{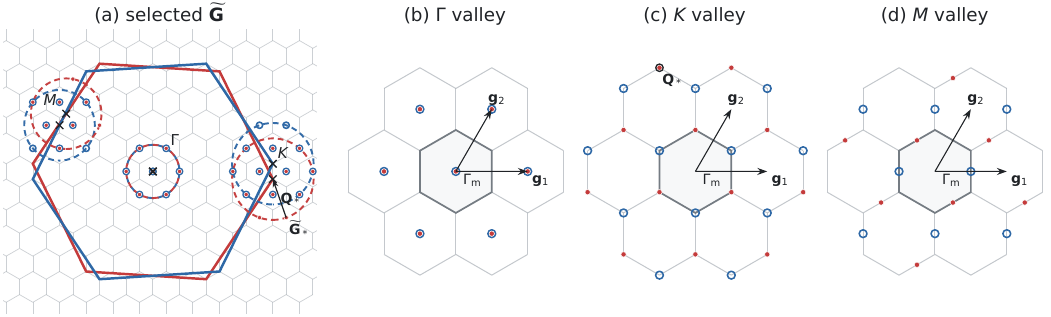}
  \caption{TAPW reciprocal-space construction for representative \(\Gamma\), \(K\), and \(M\) valleys. The schematic uses the commensurate integer pair \((m,n)=(4,5)\), corresponding to \(\theta=7.34^\circ\). (a) Reciprocal vectors \(\widetilde{\mathbf G}\) included in the TAPW basis and valley centers \(\mathbf K_\varsigma\); dashed circles indicate complete radial shells ordered by \(\lvert\mathbf Q\rvert\). The gray hexagon is the moiré Brillouin zone, and the large red and blue hexagons are the first Brillouin zones associated with the two twist groups. (b)--(d) The corresponding sets \(\mathcal Q_\varsigma\) defined by Eq.~\eqref{eq:tapw_Q_definition}. Filled red and open blue circles distinguish the two twist groups. The arrow in (a) and the thin circle in (c) indicate the same representative vector \(\mathbf Q_*\).
    }
  \label{fig:tapw_valley_selection}
\end{figure*}

For a moiré reciprocal vector \(\widetilde{\bG}\) included in the TAPW basis, define \(n=(\ell,\widetilde{\bG},\alpha)\). Let \(N_{\rm a}\) denote the number of monolayer unit cells of layer \(\ell\) within the reference moiré cell. The corresponding truncated atomic plane wave ~\cite{miao_truncated_2023} is
\begin{equation}
  \begin{aligned}
    |\psi_n^{\rm TAPW}(\kvec)\rangle
    &=
    \frac{1}{\sqrt{N_{\rm m}N_{\rm a}}}
    \sum_{I,i}
    e^{i(\kvec+\widetilde{\bG})\cdot
    (\bR_I+\bm\tau_{\ell i\mu})}
    |\bR_I;\ell,i,\alpha\rangle\\
    &=
    \sum_p|\chi_p(\kvec)\rangle X_{pn}.
  \end{aligned}
  \label{eq:tapw_basis}
\end{equation}
Here \(X\) is the \(\kvec\)-independent TAPW projection matrix, with \(X_{(\ell'i\alpha'),(\ell\widetilde{\bG}\alpha)} =\delta_{\ell'\ell}\delta_{\alpha'\alpha} e^{i\widetilde{\bG}\cdot\bm\tau_{\ell i\mu}}/\sqrt{N_{\rm a}}\) which satisfies \(X^\dagger X=I\), as derived in Appendix~\ref{app:tapw_source_symmetry_details}.

The TAPW Hamiltonian and overlap matrix are then
\begin{equation}
  \begin{aligned}
    \Ht(\kvec)
    &=X^\dagger H^{\rm loc}(\kvec)X,\\
    \St(\kvec)
    &=X^\dagger S^{\rm loc}(\kvec)X.
  \end{aligned}
\end{equation}
L\"owdin orthogonalization~\cite{lowdin_NonOrthogonality_1950} gives
\begin{equation}
  \Horth(\kvec)
  =
  \St(\kvec)^{-1/2}
  \Ht(\kvec)
  \St(\kvec)^{-1/2}.
  \label{eq:orthogonal_source_hamiltonian}
\end{equation}
All projections used to select the low-energy states below are carried out in this orthonormalized TAPW coefficient space.

A twist group \(s\) contains the layers that are rotated together. A valley \(\nu\) in twist group \(s\) defines a valley--twist sector \(\varsigma=(\nu,s)\), with valley center \(\mathbf K_\varsigma\equiv\mathbf K_{\nu,s}\). For example, in A--AB trilayer MoTe$_2$, the \(A_1\) monolayer forms one twist group and the \(A_2B\) bilayer forms the other ~\cite{qi_ChernSelective_2025}.

Let \(\widetilde{\mathcal G}_\varsigma\) denote the moiré reciprocal vectors included in the truncated TAPW basis for sector \(\varsigma\). For each \(\widetilde{\bG}\in\widetilde{\mathcal G}_\varsigma\), define
\begin{equation}
  \bQ=\mathbf K_\varsigma-\widetilde{\bG},
  \qquad
  \mathcal Q_\varsigma
  =
  \left\{
    \mathbf K_\varsigma-\widetilde{\bG}
    \,\middle|\,
    \widetilde{\bG}\in\widetilde{\mathcal G}_\varsigma
  \right\}.
  \label{eq:tapw_Q_definition}
\end{equation}
The \(\bQ\) notation organizes the low-energy states and the continuum basis: \(\kvec+\widetilde{\bG}=\mathbf K_\varsigma+(\kvec-\bQ)\), so \(\kvec-\bQ\) is the momentum measured from the sector's valley center. The TAPW cutoff \(n_g\) selects complete radial shells ordered by \(\lvert\bQ\rvert\), thereby fixing \(\widetilde{\mathcal G}_\varsigma\) and \(\mathcal Q_\varsigma\). Figure~\ref{fig:tapw_valley_selection} illustrates the construction for representative \(\Gamma\), \(K\), and \(M\) valleys. Further details on the sector labels and shared \(\bQ\) sets are given in Appendix~\ref{app:tapw_source_symmetry_details}.

For a unitary or antiunitary symmetry \(g\), the atom maps and orbital--spin rotations defined above give the matrix \(D_g^{\rm loc}\). Let \(\eta_g=0\) for a unitary operation and \(\eta_g=1\) for an antiunitary operation. Its action on the localized-orbital Bloch basis is
\begin{equation}
  \widehat g\,|\chi_p(\kvec)\rangle
  =
  \sum_{p'}|\chi_{p'}(\mathcal R_g\kvec)\rangle
  [D_g^{\rm loc}]_{p'p},
  \label{eq:localized_bloch_symmetry_action}
\end{equation}
where \(\mathsf O_g\) is the real-space point-operation matrix and \(\mathcal R_g=(-1)^{\eta_g}\mathsf O_g^{-T}\). An antiunitary operation additionally complex conjugates the expansion coefficients. Thus its action on Bloch coefficient vectors is \(D_g^{\rm loc}\mathcal K^{\eta_g}\), where \(\mathcal K\) denotes complex conjugation. We use \(M^{[0]}=M\) and \(M^{[1]}=M^*\).

Each candidate real-space symmetry \(g\) maps a TAPW channel as \((\varsigma,\bQ)\mapsto(g\varsigma,Q_g(\varsigma,\bQ))\), where \(Q_g(\varsigma,\bQ)=\mathcal R_g\bQ\). The operation can be represented within the selected TAPW basis only if this map permutes the selected channels. Operations that map to a valley outside the selected sectors relate different valley models; they are not imposed as internal symmetries of the selected model.

For an operation that preserves the selected TAPW space, the localized-orbital action satisfies the intertwining relation below. Since \(X^\dagger X=I\), its TAPW representation is
\begin{equation}
  \begin{aligned}
    D_g^{\rm loc}X^{[\eta_g]}
    &=XD_g^{\rm TAPW},\\
    D_g^{\rm TAPW}
    &=X^\dagger D_g^{\rm loc}X^{[\eta_g]}.
  \end{aligned}
  \label{eq:localized_to_tapw_symmetry}
\end{equation}
The atom map, real-space block transformation, reciprocal-label map, and Brillouin-zone-folding phase are given in Appendix~\ref{app:tapw_source_symmetry_details}. The full TAPW action is \(\mathsf S_g^{\rm TAPW}=D_g^{\rm TAPW}\mathcal K^{\eta_g}\). When both \(\Ht\) and \(\St\) obey covariance under this action, L\"owdin orthogonalization preserves the same representation (Appendix~\ref{app:subsec:source_covariance}), so
\begin{equation}
  \Horth(\kvec)
  =
  D_g^{\rm TAPW}
  \left[
    \Horth(\mathcal R_g^{-1}\kvec)
  \right]^{[\eta_g]}
  D_g^{{\rm TAPW}\dagger}.
  \label{eq:saved_source_raw_covariance}
\end{equation}

Using \(\widetilde{\bG}=\mathbf K_\varsigma-\bQ\), we relabel the TAPW basis as \(\lvert\varsigma,\bQ,c\rangle\), where \(c=(\ell,\alpha)\) collects the physical-layer, atomic-orbital, and spin labels. In this basis,
\begin{equation}
  \left[D_g^{\rm TAPW}\right]_
  {(\varsigma',\bQ',c'),(\varsigma,\bQ,c)}
  =
  \delta_{\varsigma',g\varsigma}\,
  \delta_{\bQ',Q_g(\varsigma,\bQ)}
  \left[d_{g,\varsigma}\right]_{c'c}.
  \label{eq:source_fiber_route_main}
\end{equation}
The Kronecker deltas specify the sector and plane-wave channel maps. In the phase convention used here, \(d_{g,\varsigma}\) contains the internal layer, orbital, and spin action together with the valley-folding phase, and is independent of \(\bQ\) (Appendix~\ref{app:subsec:local_orbital_symmetry}).

\subsection{Low-energy basis and effective Hamiltonian}\label{sec:low_energy_basis}

\begin{figure*}[!t]
  \centering
  \includegraphics[width=0.98\textwidth]
  {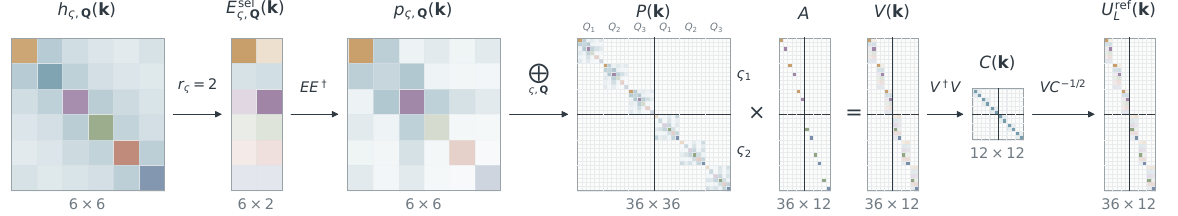}
  \caption{Low-energy projection and construction of the reference-state basis. For a \(\bQ\) block that belongs to one sector, the selected eigenvectors define \(p_{\varsigma,\bQ}\). The direct sum of these projectors gives \(P\). Projecting the reference-state matrix gives \(V=PA\), which is orthonormalized to obtain \(U_L^{\rm ref}=V(V^\dagger V)^{-1/2}\).}
  \label{fig:low_energy_basis_construction}
\end{figure*}

Following Ref.~\cite{zhang_universal_2024}, we select the low-energy states by diagonalizing the corresponding block of \(\Horth(\kvec)\) at each fixed \(\bQ\). The equations below assume that each fixed-\(\bQ\) block belongs to a single valley--twist sector \(\varsigma\); we denote this block by \(h_{\varsigma,\bQ}(\kvec)\). The extension to jointly selected sectors, as in the \(\Gamma\)-valley models, is given in Appendix~\ref{app:subsec:reference_orbital_basis}.

For each sector \(\varsigma\), we retain the same number \(r_\varsigma\) of states at every \(\bQ\in\mathcal Q_\varsigma\) and at every construction momentum. Their orthonormal eigenvectors define
\begin{equation}
  \begin{aligned}
    h_{\varsigma,\bQ}(\kvec)E_{\varsigma,\bQ}^{\rm sel}(\kvec)
    &=E_{\varsigma,\bQ}^{\rm sel}(\kvec)
      \varepsilon_{\varsigma,\bQ}^{\rm sel}(\kvec),\\
    p_{\varsigma,\bQ}(\kvec)
    &=E_{\varsigma,\bQ}^{\rm sel}(\kvec)
      E_{\varsigma,\bQ}^{{\rm sel}\dagger}(\kvec),\\
    P(\kvec)
    &=\bigoplus_{\substack{\varsigma\\
                           \bQ\in\mathcal Q_\varsigma}}
      p_{\varsigma,\bQ}(\kvec).
  \end{aligned}
  \label{eq:qblock_low_projector}
\end{equation}
Here \(p_{\varsigma,\bQ}(\kvec)\) projects onto the \(r_\varsigma\)-dimensional low-energy subspace selected from the fixed-\(\bQ\) block \(h_{\varsigma,\bQ}(\kvec)\).  The selected states from all fixed-\(\bQ\) blocks together span the low-energy subspace, whose projector is \(P(\kvec)\). The couplings between different \(\bQ\) channels are not used to define this initial projector, but they remain in \(\Horth(\kvec)\) and are included in the construction of \(H_{\rm eff}\) below.

Numerical diagonalization determines the selected low-energy subspace but does not define a unique basis within it. An isolated eigenstate is defined only up to a phase, and a degenerate multiplet admits unitary rotations among its eigenvectors. To assign layer, orbital, and spin labels, \moirekp{} projects reference vectors associated with localized-orbital channels or specified linear combinations of those channels. These vectors are represented in the orthonormalized TAPW coordinates and are chosen to be mutually orthonormal before projection. Column-pivoted QR of the projected candidate matrix identifies a linearly independent set, following the column-selection idea of the SCDM construction~\cite{damle_scdm_2015,damle_scdmk_2017}. For the symmetry construction below, this choice must also satisfy the reference-space covariance condition in Eq.~\eqref{app:eq:projector_and_reference_covariance}.

The ordered reference states define the internal labels \(\xi=1,\ldots,r_\varsigma\) within each sector. The continuum plane-wave basis is
\begin{equation}
  \begin{aligned}
    \gamma&=(\varsigma,\xi,\bQ),\quad
    \bQ\in\mathcal Q_\varsigma,\quad
    \xi=1,\ldots,r_\varsigma,\\
    \mathcal B_{\rm cont}(\kvec)
    &=
    \left\{
      |\kvec;\varsigma,\xi,\bQ\rangle
    \right\}_{\varsigma,\xi,\bQ},
    \quad
    N_{\rm cont}
    =\sum_\varsigma r_\varsigma|\mathcal Q_\varsigma|.
  \end{aligned}
  \label{eq:continuum_composite_index}
\end{equation}
The internal state \(\xi\) may be a linear combination of layer, orbital, and spin states. We use \(\gamma\) as a compact label for the continuum basis.

Let \(A\in\mathbb C^{N_{\rm TAPW}\times N_{\rm cont}}\) collect the ordered reference states associated with all continuum-basis labels \(\gamma=(\varsigma,\xi,\bQ)\), expressed in the same orthonormalized TAPW coordinates as \(P\), with \(A^\dagger A=I\). Their projection into the selected low-energy subspace and subsequent orthonormalization give
\begin{equation}
  \begin{aligned}
    V(\kvec)&=P(\kvec)A,\\
    C(\kvec)&=V^\dagger(\kvec)V(\kvec),\\
    U_L^{\rm ref}(\kvec)
    &=V(\kvec)C^{-1/2}(\kvec).
  \end{aligned}
  \label{eq:embedded_coordinate_anchors}
\end{equation}
Equation~\eqref{eq:embedded_coordinate_anchors} requires \(C(\kvec)=A^\dagger P(\kvec)A\) to be positive definite at each momentum where the basis is used. Then \(U_L^{{\rm ref}\dagger}U_L^{\rm ref}=I\) and \(U_L^{\rm ref}U_L^{{\rm ref}\dagger}=P\): the reference-state construction fixes a basis without changing the selected subspace. The same reference-state pattern and ordering are used at every momentum and in every \(\bQ\) channel of a sector. Figure~\ref{fig:low_energy_basis_construction} summarizes this construction.
Appendix~\ref{app:subsec:reference_diagnostics} reports the minimum reference-overlap eigenvalues over all retained \(Q\) channels and construction momenta, together with the reference-space covariance checks.

Although \(P(\kvec)\) is constructed from the fixed-\(\bQ\) blocks, the full Hamiltonian \(\Horth(\kvec)\) contains couplings between different \(\bQ\) channels and between the selected and excluded states. Let \(U_H(\kvec)\) be an orthonormal basis for the complement of the selected low-energy subspace. With \(U_L(\kvec)=U_L^{\rm ref}(\kvec)\), define \(H_{XY}^{\rm ref}(\kvec)=U_X^\dagger(\kvec)\Horth(\kvec)U_Y(\kvec)\) for \(X,Y\in\{L,H\}\). The fixed-reference-energy form of L\"owdin partitioning~\cite{lowdin_Note_1951} is
\begin{equation}
  \begin{aligned}
    H_{\rm eff}^{\rm ref}(\kvec;E_{\rm ref})
    &=
    H_{LL}^{\rm ref}(\kvec)\\
    &\;+\;
    H_{LH}^{\rm ref}(\kvec)
    \left[
      E_{\rm ref}I-H_{HH}^{\rm ref}(\kvec)
    \right]^{-1}
    H_{HL}^{\rm ref}(\kvec).
  \end{aligned}
  \label{eq:projected_heff}
\end{equation}
The first term contains the full Hamiltonian within the selected subspace, including couplings between different \(\bQ\) channels. The second accounts for virtual mixing with the excluded states. The user specifies one reference energy \(E_{\rm ref}\) per model, held fixed throughout the construction. We retain the first-order energy dependence of the eliminated-space correction. With \(R(\kvec)=[E_{\rm ref}I-H_{HH}^{\rm ref}(\kvec)]^{-1}\) and \(D(\kvec)=H_{LH}^{\rm ref}(\kvec)R^2(\kvec)H_{HL}^{\rm ref}(\kvec)\), the eliminated components contribute an effective overlap matrix \(S_{\rm eff}(\kvec)=I+D(\kvec)\). Orthogonalization with its inverse square root gives the static Hamiltonian~\cite{solovyev_Construction_2007}
\begin{equation}
  \begin{aligned}
    H_{\rm eff}^{\rm lin}(\kvec)
    &=S_{\rm eff}^{-1/2}(\kvec)\Bigl[H_{LL}^{\rm ref}(\kvec)\\
    &\quad+H_{LH}^{\rm ref}(\kvec)R(\kvec)H_{HL}^{\rm ref}(\kvec)\\
    &\quad+E_{\rm ref}D(\kvec)\Bigr]S_{\rm eff}^{-1/2}(\kvec).
  \end{aligned}
  \label{eq:projected_heff_linearized}
\end{equation}
All six models use Eq.~\eqref{eq:projected_heff_linearized}, denoted by \(H_{\rm eff}\) below. Appendix~\ref{app:subsec:downfold_sensitivity} gives its derivation and comparison with Eq.~\eqref{eq:projected_heff}, together with the target windows, separation from excluded states, and reference-energy sensitivity. Any fixed rotation used to order the internal states is absorbed into this basis convention and applied identically in every \(\bQ\) channel.

The overlaps in Figs.~\ref{fig:mote2_k_flagship}--\ref{fig:mote2_aab_multivalley} use the low-space embedding defined in Sec.~\ref{sec:validation_protocol}, without reconstructing the eliminated components.

\moirekp{} can automatically select the low-energy dimension by enumerating complete multiplets from the target band edge within a specified candidate sequence. The sector-dependent acceptance and tie-breaking rules are given in Appendix~\ref{app:subsec:paper_model_records}. Reference-frame symmetry is checked separately after selection. The search raises an error if no candidate satisfies the required criteria. Alternatively, the selected states and reference vectors can be specified by the user.

\subsection{Continuum symmetry representation}\label{sec:continuum_symmetry_transfer}

\(D_g^{\rm TAPW}\) acts in the full TAPW space, whereas the continuum model contains only the selected low-energy states. For the \(j\)th symmetry-related momentum pair \((\mathcal R_g^{-1}\kvec_j,\kvec_j)\), projection onto the corresponding reference-state bases gives the sewing matrix
\begin{equation}
  B_{g,j}^{\rm low}
  =
  U_L^{{\rm ref}\dagger}(\kvec_j)\,
  D_g^{\rm TAPW}\,
  \left[U_L^{\rm ref}(\mathcal R_g^{-1}\kvec_j)\right]^{[\eta_g]}.
  \label{eq:projected_low_action}
\end{equation}
When both the selected projectors and the ordered reference subspaces obey the covariance relations in Eq.~\eqref{app:eq:projector_and_reference_covariance}, the sector-resolved projected action separates into a \(\bQ\)-channel map and an internal matrix:
\begin{equation}
  \left[B_{g,j}^{\rm low}\right]_{
    (\varsigma',\xi',\bQ'),(\varsigma,\xi,\bQ)}
  =
  \delta_{\varsigma',\,g\varsigma}
  \delta_{\bQ',\,Q_g(\varsigma,\bQ)}
  [\rho_{g,\varsigma}]_{\xi'\xi}.
  \label{eq:q_uniform_projected_action}
\end{equation}
The sector and \(\bQ\)-channel maps are the same as in the TAPW representation of Eq.~\eqref{eq:source_fiber_route_main}; the low-energy projection replaces \(d_{g,\varsigma}\) by \(\rho_{g,\varsigma}\). Its independence of \(\kvec\) and \(\bQ\) follows from both covariance conditions and the momentum-independent reference pattern. The derivation is given in Appendix~\ref{app:low_energy_symmetry}.

\(H_{\rm eff}\) and the projected symmetry matrices are expressed in the same low-energy basis. For each generator, one specified symmetry-related momentum pair provides \(\rho_{g,\varsigma}\). Additional momentum pairs check subspace leakage, projected-action unitarity, and covariance of \(H_{\rm eff}\). After these checks, small residual deviations within the selected internal blocks are removed by numerically enforcing unitarity and the required group relations. This correction changes neither the selected low-energy subspace nor the sector and \(\bQ\)-channel maps, and cannot repair appreciable leakage. We denote the corrected internal matrices by \(\rho_{g,\varsigma}^{\rm cont}\).

Combining \(\rho_{g,\varsigma}^{\rm cont}\) with the sector and \(\bQ\)-channel maps gives the continuum symmetry representation \(\mathsf S_g^{\rm cont}=U_g^{\rm cont}\mathcal K^{\eta_g}\):
\begin{equation}
  \begin{aligned}
    \relax[U_g^{\rm cont}]_{\gamma'\gamma}
    &=
    [\rho_{g,\varsigma}^{\rm cont}]_{\xi'\xi}\,
    \delta_{\varsigma',g\varsigma}
    \delta_{\bQ',Q_g(\varsigma,\bQ)},\\
    H_{\rm cont}(\kvec)
    &=
    U_g^{\rm cont}
    \left[
      H_{\rm cont}\!\left(\kappa_g(\kvec)\right)
    \right]^{[\eta_g]}
    U_g^{{\rm cont}\dagger}.
  \end{aligned}
  \label{eq:continuum_action_matrix}
\end{equation}
Here \(\kappa_g(\kvec)=\mathcal R_g^{-1}\kvec\). The composite indices \(\gamma\) and \(\gamma'\) follow the convention in Eq.~\eqref{eq:continuum_composite_index}. The second line of Eq.~\eqref{eq:continuum_action_matrix} is imposed by the symmetry-allowed operator construction in Sec.~\ref{sec:allowed_hamiltonian_terms}. For the six models, we separately checked the covariance of \(H_{\rm eff}\) under the corrected action. Appendix~\ref{app:subsec:continuum_sewing} gives the residual definitions, tolerance and results.

\subsection{Symmetry-allowed Hamiltonian terms and their coefficients}\label{sec:allowed_hamiltonian_terms}

A continuum Hamiltonian in the plane-wave basis contains momentum dispersion and internal-state mixing within each \(\bQ\) channel, moiré scattering between different \(\bQ\) channels, and coupling between different valley--twist sectors. The selected sets \(\mathcal Q_\varsigma\) fix the plane-wave basis. Within this basis, a truncation \(\mathfrak M\) specifies the internal-state connections, moiré momentum-transfer harmonics, and maximum polynomial degree retained in the continuum Hamiltonian.

When the numbers of intralayer and interlayer harmonics are not specified, \moirekp{} first compares harmonic truncations of \(H_{\rm eff}\), as in Eq.~\eqref{app:eq:harmonic_ablation}. It selects the candidate with the fewest shells that satisfies the stated band-error and low-energy-subspace-overlap criteria within the candidate set. The maximum polynomial degree is held fixed. The errors of the final \(H_{\rm cont}\) are evaluated after coefficient extraction. Appendix~\ref{app:model_parameter_determination} gives the candidate shells, selection criterion, and comparison momenta.

Following the sector-resolved plane-wave construction of Ref.~\cite{zhang_universal_2024}, let \(a=(\varsigma_a,\xi_a)\) and \(b=(\varsigma_b,\xi_b)\) denote the internal row and column states. A moiré Fourier harmonic connects them with momentum transfer
\[
  \mathbf p
  =
  \mathbf K_{\varsigma_a}
  -\mathbf K_{\varsigma_b}
  +\bG_M,
\]
where \(\bG_M\) is a moiré reciprocal-lattice vector. For \(\bQ_1\in\mathcal Q_{\varsigma_a}\) and \(\bQ_2\in\mathcal Q_{\varsigma_b}\), we define
\begin{equation}
  \begin{aligned}
  &\left[Y_{ab,\mathbf p}^{mn}(\kvec)\right]_
  {(a_1,\bQ_1),(a_2,\bQ_2)}\\
  &\quad=
  \delta_{a_1,a}\,
  \delta_{a_2,b}\,
  \delta_{\bQ_1,\bQ_2+\mathbf p}\,
  z_{\bQ_1}(\kvec)^m
  \bar z_{\bQ_1}(\kvec)^n,
  \end{aligned}
  \label{eq:raw_elementary_continuum_term}
\end{equation}
where \(z_{\bQ}(\kvec)=(k_x-Q_x)+i(k_y-Q_y)\) and \(\bar z_{\bQ}(\kvec)=z_{\bQ}(\kvec)^*\). Here \(m,n\) are nonnegative integers, and \(k_x,k_y,Q_x,Q_y\) are Cartesian components. The total polynomial degree is \(m+n\). The elementary term is generally neither Hermitian nor symmetry invariant.

The choice \((m,n)=(0,0)\) gives a momentum-independent term; \((1,0)\) and \((0,1)\) give the linear \(z\) and \(\bar z\) terms; \((1,1)\) gives \(|\kvec-\bQ_1|^2\); and \((2,0)\) and \((0,2)\) give quadratic terms with opposite angular dependence. Together with \((a,b)\) and \(\mathbf p\), these monomials describe momentum dispersion, onsite energies, internal-state mixing, intralayer moiré potentials, and interlayer coupling.

The continuum symmetry representation acts on each elementary term as
\begin{equation}
  \mathcal T_g
  \left[Y_{ab,\mathbf p}^{mn}\right](\kvec)
  =
  U_g^{\rm cont}
  \left[
    Y_{ab,\mathbf p}^{mn}\!
    \left(\kappa_g(\kvec)\right)
  \right]^{[\eta_g]}
  U_g^{{\rm cont}\dagger}.
  \label{eq:continuum_term_group_action}
\end{equation}
Starting from the elementary terms included in \(\mathfrak M\), \moirekp{} forms Hermitian combinations invariant under the selected symmetry operations. Both \(Y\) and \(iY\) are included as real seed directions, as described in Appendix~\ref{app:symmetry_allowed_terms}. The seeds need not themselves be closed under those operations: their symmetry images and adjoints define the completed candidate space. Its Hermitian invariant part is a finite real vector space, denoted by \(\mathcal V_{\rm sym}(\mathfrak M)\). We denote an independent basis of this space by \(\{\widetilde Y_j^{(\mathfrak M)}(\kvec)\}\). Appendix~\ref{app:symmetry_allowed_terms} gives the elementary-term construction, closure, symmetry projection, and removal of linear dependencies.

We next determine the coefficients of these operators from \(\heff(\kvec)\). We orthonormalize the independent allowed operators with respect to the matrix inner product \(\langle\cdot,\cdot\rangle_{\mathfrak M}\) and denote the resulting basis by \(\{Y_j^{(\mathfrak M)}(\kvec)\}\). Following Ref.~\cite{zhang_universal_2024}, the coefficients are
\begin{equation}
  \begin{aligned}
  \theta_j
  &=
  \left\langle
    Y_j^{(\mathfrak M)},\heff
  \right\rangle_{\mathfrak M},\\
  H_{\rm cont}^{(\mathfrak M)}(\kvec)
  &=
  \sum_{j=1}^{N_{\rm term}(\mathfrak M)}
  \theta_jY_j^{(\mathfrak M)}(\kvec).
  \end{aligned}
  \label{eq:trace_projection_coefficients}
\end{equation}
The coefficients are determined from Hamiltonian matrix elements rather than band energies.

\subsection{Band and wave-function comparisons}\label{sec:validation_protocol}

After selecting the truncation, let \(\Hcont(\kvec)=H_{\rm cont}^{(\mathfrak M)}(\kvec)\) denote the final continuum Hamiltonian. We separately evaluate the TAPW--\(H_{\rm eff}\), \(H_{\rm eff}\)--\(H_{\rm cont}\), and TAPW--\(H_{\rm cont}\) RMS energy differences within the target-band windows along the high-symmetry \(k\) paths. The windows and results are listed in Table~\ref{tab:paper_model_configurations}. These errors use the energy alignment of Figs.~\ref{fig:mote2_k_flagship}--\ref{fig:mote2_aab_multivalley}, with the \(H_{\rm eff}\) alignment shown in Fig.~\ref{fig:downfold_reference_sensitivity}.

For the overlaps in Figs.~\ref{fig:mote2_k_flagship}--\ref{fig:mote2_aab_multivalley}, a normalized continuum eigenvector is embedded in the orthonormal TAPW coefficient space as \(U_L(\kvec)v_n^{\rm cont}(\kvec)\). For an isolated matched state, the overlap is
\[
 \mathcal O_n(\kvec)=\left|\left\langle\psi_{m(n,\kvec)}^{\rm TAPW}\middle|U_L(\kvec)v_n^{\rm cont}\right\rangle\right|^2.
\]
No eliminated-state reconstruction or renormalization by projected TAPW weight is applied. The one-to-one band assignment \(m(n,\kvec)\) is made within a fixed TAPW target set; the matching choices are specified in Appendix~\ref{app:subsec:paper_model_records}.

A cluster joins matched pairs connected by gaps of at most \(3\) meV between neighboring levels in either energy-ordered spectrum. Each of the \(d\) members of a cluster \(\mathcal C\) is assigned the same subspace overlap
\[
 \mathcal O_{\rm cluster}=\frac{1}{d}\left\|(\Psi_{\mathcal C}^{\rm TAPW})^\dagger U_L V_{\mathcal C}^{\rm cont}\right\|_{\rm F}^2.
\]
The 3 meV threshold groups nearby bands for evaluating these overlaps and is separate from eigensolver or model-selection tolerances.

At a high-symmetry momentum \(\kvec_\star\), let the columns of \(V_{\mathcal B}(\kvec_\star)\) be orthonormal eigenstates of \(H_{\rm cont}\) in the band set \(\mathcal B\). If \(\mathcal R_g\kvec_\star=\kvec_\star+\Delta\bG\), the projected symmetry action in this subspace is
\begin{equation}
  \mathcal D_{g,\mathcal B}(\kvec_\star)
  =
  V_{\mathcal B}^{\dagger}(\kvec_\star)
  \mathcal S_{-\Delta\bG}^{\rm mod}
  U_g^{\rm cont}
  V_{\mathcal B}^{[\eta_g]}(\kvec_\star).
  \label{eq:band_symmetry_representation}
\end{equation}
Here \(\mathcal S_{-\Delta\bG}^{\rm mod}\) is the reciprocal-boundary sewing matrix defined in Eq.~\eqref{app:eq:model_boundary_sewing}. When the combined symmetry and reciprocal shift permutes the selected plane-wave channels and preserves \(\mathcal B\), Eq.~\eqref{eq:band_symmetry_representation} defines the little-group action in this subspace. For a unitary operation, this action is linear and its eigenvalues are the symmetry eigenvalues. For an antiunitary operation, complex conjugation is part of the action, so \(\mathcal D_{g,\mathcal B}\) is its sewing matrix and its eigenvalues are not symmetry labels.

For a band subspace \(\mathcal B\) in the orthonormal model coefficient basis, let \(P_{\mathcal B}=V_{\mathcal B}V_{\mathcal B}^{\dagger}\) and \(\partial_i=\partial/\partial k_i\), where \(k_i\) denotes a Cartesian momentum component. We use the Berry connection \(\mathcal A_i=+i\langle u|\partial_i u\rangle\). The quantum metric and Berry curvature are
\begin{equation}
  \begin{aligned}
    g_{ij}
    &=
    \frac{1}{2}\operatorname{Tr}
    \left[
      \partial_iP_{\mathcal B}
      \partial_jP_{\mathcal B}
    \right],\\
    \Omega_{ij}
    &=
    +i\operatorname{Tr}
    \left\{
      P_{\mathcal B}
      \left[
        \partial_iP_{\mathcal B},
        \partial_jP_{\mathcal B}
      \right]
    \right\}.
  \end{aligned}
  \label{eq:projector_quantum_geometry}
\end{equation}
The plotted quantities are evaluated from finite differences of the band projector, with \(\operatorname{tr}g=g_{xx}+g_{yy}\).

For Figs.~\ref{fig:mote2_k_flagship}--\ref{fig:mote2_aab_multivalley}, Eq.~\eqref{eq:projector_quantum_geometry} is evaluated in the respective orthonormal coefficient representations and defines model-space geometry, not the complete microscopic Bloch-state geometry. No cross-\(\kvec\) atomic-orbital basis overlaps or eliminated-state reconstruction are included.

For the geometric maps, \(\mathbf b_M=\mathbf b_1\) is the first primitive moir\'e reciprocal vector and \(A_{\rm mBZ}=|\det(\mathbf b_1,\mathbf b_2)|\). The plotted fields are \((A_{\rm mBZ}/2\pi)\Omega_{xy}\) and \((A_{\rm mBZ}/2\pi)(g_{xx}+g_{yy})\), with Cartesian coordinates scaled by \(|\mathbf b_M|\).

Wilson-loop centers are obtained from the eigenphases of the ordered product of neighboring-subspace overlap matrices along a closed momentum-space path. Denoting this product by \(\mathcal W_{\mathcal C}\),
\[
  \nu_a=\frac{1}{2\pi}\arg w_a,
  \qquad
  w_a\in\operatorname{eig}(\mathcal W_{\mathcal C}).
\]

The Wilson loops run along increasing \(\kappa_2\), with 41 points on each fractional axis and both endpoints retained. For frames \(V_j\), \(j=0,\ldots,N-1\), the interior links are \(L_j=\operatorname{polar}(V_{j+1}^{\dagger}V_j)\), \(j=0,\ldots,N-2\), and the closing link is \(L_{N-1}=\operatorname{polar}(V_0^{\dagger}\mathcal S V_{N-1})\), where \(\mathcal S\) maps the endpoint to the starting reciprocal representative. The ordered product is \(\mathcal W_{\mathcal C}=L_{N-1}\cdots L_1L_0\). This convention gives the single-band phase \(+\oint\mathcal A_i\,dk_i\) modulo \(2\pi\). Appendix~\ref{app:subsec:continuum_sewing} defines the reciprocal-boundary map and reports cutoff and mesh comparisons.

\section*{Data availability}

The atomic structures and OpenMX input files for the examples studied in this work are available at \url{https://github.com/zhangyan31415/MoireKP}. The repository also provides model configurations and scripts for constructing the continuum models and calculating their band structures and quantum-geometric properties. Large Hamiltonian and overlap matrices can be regenerated from the supplied first-principles inputs following the instructions in the repository.

\section*{Code availability}

The MoireKP source code is available at \url{https://github.com/zhangyan31415/MoireKP} under the GNU Lesser General Public License (\texttt{LGPL-3.0-or-later}). Documentation and example workflows are included in the repository.

\section*{Acknowledgements}
We thank Jiabin Yu for helpful discussions. This work was supported by the National Key R\&D Program of China (2024YFA1408400, 2023YFA1607400), the National Natural Science Foundation of China (12274436), the Science Center of the National Natural Science Foundation of China (12188101), the Beijing Municipal Science \& Technology Commission, Administrative Commission of Zhongguancun Science Park (Z251100003625025), and the Strategic Priority Research Program (B) of the Chinese Academy of Sciences (CAS) (XDB1710000). W.~M. and X.~D. were supported by the Research Grants Council of Hong Kong through the Senior Research Fellow Scheme (Project No. SRFS2324-6S01) and by the New Cornerstone Science Foundation.

\section*{Author contributions}
Q.W. supervised the project. Y.Z. conceived and carried out the work, developed the methodology and software, and wrote the initial manuscript. H.P. contributed to discussions of continuum-model parameter determination. W.M. contributed to discussions of the TAPW formulation. All authors contributed to discussions and revision of the manuscript.

\bibliographystyle{apsrev4-2}

\clearpage
\bibliography{main}

\clearpage
\onecolumngrid
\appendix
\numberwithin{equation}{section}
\renewcommand{\thesubsection}{\thesection.\arabic{subsection}}
\makeatletter
\renewcommand{\p@subsection}{}
\makeatother

\addtocontents{toc}{\protect\let\protect\contentsline\protect\AppendixContentsLine}
\begingroup
\hypersetup{linkcolor=black}
\renewcommand{\tocname}{Appendix contents}
\makeatletter
\let\AppendixContentsLine\contentsline
\let\contentsline\@gobblefour
\let\l@subsubsection\@gobbletwo
\makeatother
\tableofcontents
\endgroup
\clearpage

\section{TAPW construction and symmetry}
\label{app:tapw_source_symmetry_details}

The TAPW construction follows Refs.~\cite{miao_truncated_2023,zhang_universal_2024}. This appendix derives how a crystal symmetry is represented in the TAPW basis.  The construction has three ingredients: the Fourier embedding of localized orbitals, the atom--orbital--spin action of the symmetry, and the Brillouin-zone folding associated with the chosen valley representatives.  Their combination produces a permutation of the selected \(Q\) channels and a small internal matrix that is repeated for every allowed channel map.

\subsection{Plane-wave basis}

Physical layers are divided into twist groups \(\mathcal L_s\). For the calculations considered here, a single target valley is selected, with one valley center for each twist group. A valley--twist sector and its internal channel are
\begin{equation}
  \varsigma=(\nu,s),
  \qquad
  c=(\ell,\mu,\zeta),\quad \ell\in\mathcal L_s ,
  \label{app:eq:source_sector_and_internal_channel}
\end{equation}
where \(\nu\) is the selected valley, \(\mu\) labels a basis atom, and \(\zeta\) labels its orbital and spin component. For a chosen moiré reciprocal representative \(\widetilde{\bG}\), the corresponding plane-wave channel is
\begin{equation}
  \bQ=\mathbf K_\varsigma-\widetilde{\bG},
  \qquad
  \mathcal Q_\varsigma
  =
  \{\mathbf K_\varsigma-\widetilde{\bG}:
  \widetilde{\bG}\in\widetilde{\mathcal G}_\varsigma\}.
  \label{app:eq:group_tagged_Q_set}
\end{equation}

The full TAPW basis label is \((\varsigma,\bQ,c)\). The valley label does not introduce an additional copy of a basis state. If two twist groups have the same geometric \(\bQ\), as at \(\Gamma\), their states remain distinct because their internal coordinates belong to disjoint physical-layer subspaces.

Following Ref.~\cite{zhang_universal_2024}, sectors whose valley centers differ by a moiré reciprocal-lattice vector belong to the same subvalley, and their \(\mathcal Q\) sets lie on the same translated moiré reciprocal lattice. At \(\Gamma\), all twist groups share the same geometric \(\bQ\) set and enter a common fixed-\(\bQ\) block. Otherwise, the shifted \(\mathcal Q\) sets are treated separately.

For a commensurate structure, the common supercell basis \(T\), layer primitive bases \(A_s\), and reciprocal bases \(G\) and \(B_s\) obey
\begin{equation}
  T=A_sS_s,\qquad
  G=2\pi T^{-T},\qquad
  B_s=2\pi A_s^{-T}=GS_s^T ,
  \label{app:eq:commensurate_summary}
\end{equation}
with integer matrices \(S_s\).  Thus every layer reciprocal vector is also a moiré reciprocal vector.

Let \(\bm\rho_{\ell i}\) enumerate the primitive cells of layer \(\ell\) inside the commensurate supercell and let \(\mathbf d_{\ell\mu}\) be the intracell center of atom \(\mu\).  The reference positions used by the TAPW Fourier labels are
\begin{equation}
  \bm\tau_{\ell i\mu}^{\rm ref}
  =\bm\rho_{\ell i}+\mathbf d_{\ell\mu}.
  \label{app:eq:two_scale_atomic_center}
\end{equation}
The same rigid reference position is denoted by \(\bm\tau_{\ell i\mu}\) in the main text; the superscript \({\rm ref}\) is retained in this appendix to distinguish it from the relaxed atomic coordinates entering the OpenMX calculation. The TAPW projector and symmetry labels use this rigid commensurate reference lattice, whereas relaxation enters through the first-principles Hamiltonian and overlap matrices.  Compatibility of those matrices with each selected symmetry is checked explicitly below.

In one internal channel, the Fourier embedding is
\begin{equation}
  X_{(\ell'i\mu'\zeta'),(\varsigma,\widetilde{\bG},c)}
  =
  \frac{
    \delta_{\ell'\ell}\delta_{\mu'\mu}\delta_{\zeta'\zeta}
  }{\sqrt{N_s}}\,
  e^{i\widetilde{\bG}\cdot\bm\tau_{\ell i\mu}^{\rm ref}} .
  \label{app:eq:X_channel_normalization}
\end{equation}
Here \(N_s\) is the number of primitive cells in the translation orbit of twist group \(s\). Different twist groups have disjoint physical-layer support. Within one twist group, complete translation orbits give the finite-character orthogonality relation
\begin{equation}
  \frac{1}{N_s}\sum_i
  e^{i(\widetilde{\bG}'-\widetilde{\bG})\cdot\bm\rho_{\ell i}}
  =
  \delta_{[\widetilde{\bG}'],[\widetilde{\bG}]} ,
  \label{app:eq:finite_character_orthogonality}
\end{equation}
where the equivalence classes are taken modulo the layer reciprocal lattice. The program includes at most one representative from each class in a given twist group. Consequently
\begin{equation}
  X^\dagger X=I .
  \label{app:eq:X_isometry}
\end{equation}
The localized-orbital matrices are compressed as \(A^{\rm TAPW}=X^\dagger A^{\rm loc}X\), with \(A=H,S\).  This congruence preserves Hermiticity, and \(S^{\rm TAPW}\) remains positive definite when \(S^{\rm loc}>0\).

\subsection{Localized-orbital action and TAPW projection}
\label{app:subsec:local_orbital_symmetry}

Let \(u=(\ell,i,\mu)\) label an atomic center in the reference supercell. A unitary or antiunitary point operation, represented about the chosen origin, is written
\begin{equation}
  g=\{\mathsf O_g|\mathbf 0\}\mathcal K^{\eta_g},
  \qquad
  \mathsf O_g\bm\tau_u^{\rm ref}
  =\bm\tau_{gu}^{\rm ref}+\mathbf s_{g,u},
  \quad \mathbf s_{g,u}\in\Lambda_{\rm m}.
  \label{app:eq:atomic_center_map}
\end{equation}
All spatial operations used in the present models have zero attached translation in this convention. A nonsymmorphic operation would require an additional Bloch phase and is outside the representation written here. The matrix \(U_{g,u}\) rotates the localized orbital and spin components on that center.  In real space,
\begin{equation}
  \widehat g|\bR,u,\zeta\rangle
  =
  \sum_{\zeta'}
  |\mathsf O_g\bR+\mathbf s_{g,u},gu,\zeta'\rangle
  [U_{g,u}]_{\zeta'\zeta}.
  \label{app:eq:localized_orbital_action}
\end{equation}
For an antiunitary operation, expansion coefficients are also complex conjugated. For a real-space block connecting centers \(u\) and \(v\), this action gives
\begin{equation}
  \begin{aligned}
    \bR_g
    &=\mathsf O_g\bR+\mathbf s_{g,v}-\mathbf s_{g,u},\\
    [\widehat g\,H\widehat g^\dagger]_{gu,gv}(\bR_g)
    &=U_{g,u}[H(\bR)]_{uv}^{[\eta_g]}U_{g,v}^{\dagger},\\
    [\widehat g\,S\widehat g^\dagger]_{gu,gv}(\bR_g)
    &=U_{g,u}[S(\bR)]_{uv}^{[\eta_g]}U_{g,v}^{\dagger}.
  \end{aligned}
  \label{app:eq:localized_real_space_block_action}
\end{equation}
The entries of \([H(\bR)]_{uv}\) and \([S(\bR)]_{uv}\) are the orbital--spin matrix elements for the two atomic centers. Equation~\eqref{app:eq:localized_real_space_block_action} is the block action used in the real-space average in Eq.~\eqref{eq:real_space_symmetrization}.

In the orbital-centered Bloch gauge this action gives, for both the Hamiltonian and overlap matrices,
\begin{equation}
  A^{\rm loc}(\kvec_{\rm t})
  =
  D_g^{\rm loc}
  [A^{\rm loc}(\kvec_{\rm s})]^{[\eta_g]}
  D_g^{{\rm loc}\dagger},
  \qquad
  A\in\{H,S\},
  \label{app:eq:bloch_covariance_matrix}
\end{equation}
where
\begin{equation}
  [D_g^{\rm loc}]_{(v,\zeta'),(u,\zeta)}
  =
  \delta_{v,gu}[U_{g,u}]_{\zeta'\zeta},
  \qquad
  \kvec_{\rm t}=\mathcal R_g\kvec_{\rm s},
  \quad
  \mathcal R_g=(-1)^{\eta_g}\mathsf O_g^{-T}.
  \label{app:eq:Dloc_and_momentum_action}
\end{equation}
The atomic permutation and the on-site orbital--spin rotations make \(D_g^{\rm loc}\) unitary.

The valley representative fixes the \(Q\)-channel map.  If \(\widetilde{\bG}\) belongs to sector \(\varsigma\), its image is
\begin{equation}
  \widetilde{\bG}'
  =
  \mathcal R_g(\widetilde{\bG}-\mathbf K_\varsigma)
  +\mathbf K_{g\varsigma},
  \qquad
  Q_g(\varsigma,\bQ)
  =
  \mathcal R_g\bQ .
  \label{app:eq:G_and_Q_map}
\end{equation}
MoireKP evaluates this map on every selected label and accepts the operation only if it is a bijection between the preimage and image \(Q\) sets.  This check is distinct from identifying the point-group operation in the structure.

The transformed valley center may lie outside the chosen first Brillouin zone.  The associated shift is
\begin{equation}
  \Delta\bG_{g,\varsigma}
  =
  \mathcal R_g^{-1}\mathbf K_{g\varsigma}
  -\mathbf K_\varsigma
  \in\Lambda_s^* .
  \label{app:eq:DeltaG_definition}
\end{equation}
In the orbital-centered gauge, folding it back multiplies an atom-centered orbital by
\begin{equation}
  e^{-i\Delta\bG_{g,\varsigma}\cdot
  \bm\tau_{\ell i\mu}^{\rm ref}}
  =
  e^{-i\Delta\bG_{g,\varsigma}\cdot\mathbf d_{\ell\mu}} .
  \label{app:eq:foldback_phase}
\end{equation}
The equality follows because \(\Delta\bG_{g,\varsigma}\in\Lambda_s^*\).  This gives the Brillouin-zone-folding phase in the internal symmetry matrix.  For example, \(C_{3z}\) at the \(K\) valley has \(\Delta\bG_{C_{3z},\ell}=\mathbf b_{\ell,1}\) in the convention used here; the other operations follow from Eq.~\eqref{app:eq:DeltaG_definition}.

Projecting the localized action gives
\begin{equation}
  D_g^{\rm TAPW}
  =
  X^\dagger D_g^{\rm loc}X^{[\eta_g]},
  \qquad
  D_g^{\rm loc}X^{[\eta_g]}=XD_g^{\rm TAPW}.
  \label{app:eq:Draw_projection}
\end{equation}
The second identity holds when the atomic channels and selected \(Q\) sets are closed under \(g\).  It also shows that \(D_g^{\rm TAPW}\) is unitary on the TAPW space.

The matrix element of this projection makes the plane-wave structure explicit.  Let \(n_{\rm s}=(\varsigma,\widetilde{\bG},c)\), \(n_{\rm t}=(g\varsigma,\widetilde{\bG}',c')\), and \(u_i=(\ell,i,\mu)\).  Substitution of Eq.~\eqref{app:eq:X_channel_normalization} into Eq.~\eqref{app:eq:Draw_projection} gives
\begin{equation}
  \relax[D_g^{\rm TAPW}]_{n_{\rm t}n_{\rm s}}
  =
  \frac{1}{\sqrt{N_{g\varsigma,c'}N_{\varsigma,c}}}
  \sum_i
  e^{-i\widetilde{\bG}'\cdot\bm\tau_{gu_i}^{\rm ref}}
  [U_{g,u_i}]_{\zeta'\zeta}
  e^{i(-1)^{\eta_g}\widetilde{\bG}\cdot
  \bm\tau_{u_i}^{\rm ref}} .
  \label{app:eq:Draw_matrix_element}
\end{equation}
Here \(N_{\varsigma,c}=N_s\) for a channel in twist group \(s\). It vanishes unless the sector and reciprocal labels obey \(\varsigma'=g\varsigma\) and Eq.~\eqref{app:eq:G_and_Q_map}.  For an allowed channel map, the atomic map is a bijection, so the preimage and image orbit sizes are equal.  Moreover, Eq.~\eqref{app:eq:foldback_phase} makes every term of the normalized sum identical: the part depending on the primitive-cell index is a layer reciprocal-lattice character and equals one.  The sum therefore leaves only an intracell orbital--spin matrix.

Let \(J_{\varsigma,\bQ}\) embed the internal coordinates of a fixed channel into the full TAPW space.  Combining the atom map, orbital--spin rotation, and Brillouin-zone-folding phase gives
\begin{equation}
  [d_{g,\varsigma}]_{c'c}
  =
  \begin{cases}
  \displaystyle
  e^{-i(-1)^{\eta_g}\Delta\bG_{g,\varsigma}\cdot
    \mathbf d_{\ell\mu}}\,
  [\overline U_{g,\ell\mu}]_{\zeta'\zeta},
  &
  c'=(\ell_g,\mu_g,\zeta'),\quad
  (\ell_g,\mu_g)=g(\ell,\mu),\\
  0,&\text{otherwise},
  \end{cases}
  \label{app:eq:raw_internal_block_q_independent}
\end{equation}
where \(\overline U_{g,\ell\mu}\) is the on-site orbital--spin matrix for that atomic type.  None of these factors depends on the reciprocal representative within the same sector.  Equivalently,
\begin{equation}
  [D_g^{\rm TAPW}]_
  {(\varsigma',\bQ',c'),(\varsigma,\bQ,c)}
  =
  \delta_{\varsigma',g\varsigma}
  \delta_{\bQ',Q_g(\varsigma,\bQ)}
  [d_{g,\varsigma}]_{c'c}.
  \label{app:eq:Draw_elementwise_factorization}
\end{equation}
Thus the character sum in Eq.~\eqref{app:eq:Draw_matrix_element} determines the channel map, while the surviving intracell factor determines the internal action.  In embedding notation this is
\begin{equation}
  \boxed{
  D_g^{\rm TAPW}J_{\varsigma,\bQ}^{[\eta_g]}
  =
  J_{g\varsigma,Q_g(\varsigma,\bQ)}d_{g,\varsigma},
  \qquad
  d_{g,\varsigma}\ \text{is }\bQ\text{-independent}.
  }
  \label{app:eq:raw_source_fiber_factorization}
\end{equation}
Equivalently, the full matrix is a block permutation:
\[
  D_g^{\rm TAPW}
  =
  \sum_{\varsigma,\bQ}
  J_{g\varsigma,Q_g(\varsigma,\bQ)}
  d_{g,\varsigma}
  J_{\varsigma,\bQ}^{\dagger}.
\]

\subsection{TAPW symmetry and boundary sewing}
\label{app:subsec:source_covariance}

The Brillouin-zone folding above changes the reciprocal representative within the same Bloch subspace. Reciprocal-boundary sewing instead identifies path endpoints separated by a reciprocal vector \(\mathbf b\). In the orbital-centered Bloch gauge, it is
\begin{equation}
  \begin{aligned}
    \relax[P_{\mathbf b}^{\rm loc}]_{(u,\zeta),(v,\zeta')}
    &=
    \delta_{uv}\delta_{\zeta\zeta'}
    e^{-i\mathbf b\cdot\bm\tau_u^{\rm ref}},\\
    P_{\mathbf b}^{\rm TAPW}
    &=X^\dagger P_{\mathbf b}^{\rm loc}X,\qquad
    \mathcal B_{\mathbf b}^{\rm TAPW}
    =\left(P_{\mathbf b}^{\rm TAPW}\right)^\dagger .
  \end{aligned}
  \label{app:eq:reciprocal_boundary_sewing}
\end{equation}
For the same valley--twist sector and internal channel,
\begin{equation}
  [P_{\mathbf b}^{\rm TAPW}]_
  {(\varsigma,\widetilde{\bG}',c),
   (\varsigma,\widetilde{\bG},c)}
  =
  \frac{1}{N_{\varsigma c}}
  \sum_i
  e^{-i(\widetilde{\bG}'+\mathbf b-\widetilde{\bG})
  \cdot\bm\tau_{\ell i\mu}^{\rm ref}} .
  \label{app:eq:reciprocal_boundary_shift}
\end{equation}
When \(\widetilde{\bG}'=\widetilde{\bG}-\mathbf b\) belongs to the selected set, this element equals one; a shifted column outside the set is absent. Thus the boundary-sewing matrix can be partial and need not be unitary on the full TAPW space.  By contrast, the symmetry-folding phase in Eq.~\eqref{app:eq:foldback_phase} is an internal phase under the closure conditions used to construct \(D_g^{\rm TAPW}\).

Because Eq.~\eqref{app:eq:bloch_covariance_matrix} uses the same localized-orbital action for \(H\) and \(S\), their TAPW matrices obey
\begin{equation}
  \begin{aligned}
  A_{\rm t}^{\rm TAPW}
  &=X^\dagger A_{\rm t}^{\rm loc}X\\
  &=X^\dagger D_g^{\rm loc}
    [A_{\rm s}^{\rm loc}]^{[\eta_g]}
    D_g^{{\rm loc}\dagger}X\\
  &=D_g^{\rm TAPW}
    [X^\dagger A_{\rm s}^{\rm loc}X]^{[\eta_g]}
    D_g^{{\rm TAPW}\dagger},
  \qquad A\in\{H,S\}.
  \end{aligned}
  \label{app:eq:raw_projection_chain}
\end{equation}
Here the first and last factors were replaced using the intertwining identity in Eq.~\eqref{app:eq:Draw_projection}.  Therefore
\begin{equation}
  \begin{aligned}
  H_{\rm t}^{\rm TAPW}
  &=
  D_g^{\rm TAPW}
  [H_{\rm s}^{\rm TAPW}]^{[\eta_g]}
  D_g^{{\rm TAPW}\dagger},\\
  S_{\rm t}^{\rm TAPW}
  &=
  D_g^{\rm TAPW}
  [S_{\rm s}^{\rm TAPW}]^{[\eta_g]}
  D_g^{{\rm TAPW}\dagger}.
  \end{aligned}
  \label{app:eq:common_raw_HS_action}
\end{equation}
For \(S^{\rm TAPW}>0\), functional calculus gives the same covariance for \(S^{-1/2}\).  Hence the symmetric Löwdin Hamiltonian satisfies
\begin{equation}
  \Horth(\kvec_{\rm t})
  =
  D_g^{\rm TAPW}
  [\Horth(\kvec_{\rm s})]^{[\eta_g]}
  D_g^{{\rm TAPW}\dagger}.
  \label{app:eq:lowdin_same_raw_action}
\end{equation}
No additional symmetry representation is introduced by orthogonalization. The implementation checks Eqs.~\eqref{app:eq:common_raw_HS_action} and \eqref{app:eq:lowdin_same_raw_action}.  For example, the normalized orthogonal-Hamiltonian residual is
\begin{equation}
  \epsilon_{g,{\rm TAPW}}
  =
  \frac{
    \|\Horth(\kvec_{\rm t})
    -D_g^{\rm TAPW}[\Horth(\kvec_{\rm s})]^{[\eta_g]}
     D_g^{{\rm TAPW}\dagger}\|_{\rm F}
  }{
    \|\Horth(\kvec_{\rm t})\|_{\rm F}
  }.
  \label{app:eq:saved_source_covariance_error}
\end{equation}
The \(Q\)-channel map and covariance are checked before constructing the low-energy symmetry representation.
\subsection{Real-space symmetry averaging}
\label{app:subsec:real_space_averaging_checks}

We evaluate the effect of real-space symmetry averaging using the first-principles inputs of the models in Figs.~\ref{fig:mote2_k_flagship}--\ref{fig:mote2_aab_multivalley}. The original matrices \((H^{\rm OMX},S^{\rm OMX})\) and the averaged matrices \((H^{\rm loc},S^{\rm loc})\) defined in Eq.~\eqref{eq:real_space_symmetrization} are Fourier transformed using the same convention. Their band energies are obtained in the full localized-orbital basis from
\begin{equation}
  \begin{aligned}
    H^{\rm OMX}(\kvec)c_n^{\rm OMX}(\kvec)
    &=\varepsilon_n^{\rm OMX}(\kvec)
      S^{\rm OMX}(\kvec)c_n^{\rm OMX}(\kvec),\\
    H^{\rm loc}(\kvec)c_n^{\rm loc}(\kvec)
    &=\varepsilon_n^{\rm loc}(\kvec)
      S^{\rm loc}(\kvec)c_n^{\rm loc}(\kvec).
  \end{aligned}
  \label{app:eq:averaging_band_eigenproblems}
\end{equation}
Bands are paired by energy order within each comparison window along \(\Gamma-M-K-\Gamma\). Both spectra use the original valence-band maximum as a common energy zero, without additional alignment.

For the sampled momenta \(\mathcal K\) and band set \(\mathcal I\), we define
\begin{equation}
  \begin{aligned}
    \delta\varepsilon_n(\kvec)
    &=\varepsilon_n^{\rm loc}(\kvec)
      -\varepsilon_n^{\rm OMX}(\kvec),\\
    \Delta_{\rm RMS}
    &=\left[
      \frac{1}{|\mathcal K|\,|\mathcal I|}
      \sum_{\kvec\in\mathcal K}\sum_{n\in\mathcal I}
      |\delta\varepsilon_n(\kvec)|^2
      \right]^{1/2},\\
    \Delta_{\max}
    &=\max_{\substack{\kvec\in\mathcal K\\n\in\mathcal I}}
      |\delta\varepsilon_n(\kvec)|.
  \end{aligned}
  \label{app:eq:averaging_energy_changes}
\end{equation}

\begin{figure*}[!htbp]
\centering
\includegraphics[width=0.98\textwidth]{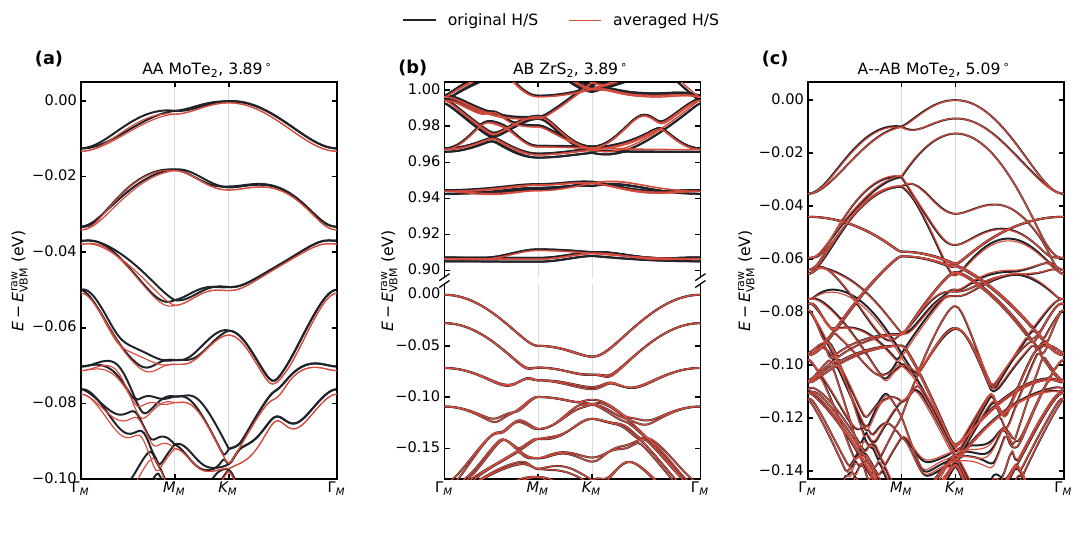}
\caption{Band structures before (black) and after (red) real-space symmetry averaging for (a) AA bilayer MoTe$_2$ at $3.89^\circ$, (b) AB bilayer ZrS$_2$ at $3.89^\circ$, and (c) A--AB trilayer MoTe$_2$ at $5.09^\circ$. Both spectra use the original VBM as the common energy zero. The valence and conduction windows in (b) are separated by an axis break.}
\label{fig:real_space_averaging_bands}
\end{figure*}

Figure~\ref{fig:real_space_averaging_bands} compares the original and averaged bands, and Table~\ref{tab:real_space_averaging_checks} reports the changes for the specified band-edge windows. For the four band-edge windows, \(\Delta_{\rm RMS}\) ranges from 0.07 to 1.19 meV and \(\Delta_{\max}\) from 0.25 to 4.59 meV.

\begin{table*}[!htbp]
\centering
\small
\caption{Band-energy changes under real-space symmetry averaging for the first-principles inputs used in this work. The comparison windows are listed explicitly; \(\Delta_{\rm RMS}\) and \(\Delta_{\max}\) are defined in Eq.~\eqref{app:eq:averaging_energy_changes}.}
\label{tab:real_space_averaging_checks}
\setlength{\tabcolsep}{7pt}
\renewcommand{\arraystretch}{1.15}
\begin{tabular}{llrr}
\toprule
Structure & Band window & $\Delta_{\rm RMS}$ (meV) & $\Delta_{\max}$ (meV) \\
\midrule
AA MoTe$_2$, $3.89^\circ$ & highest 10 valence & 1.19 & 4.59 \\
AB ZrS$_2$, $3.89^\circ$ & highest 10 valence & 0.07 & 0.25 \\
AB ZrS$_2$, $3.89^\circ$ & lowest 8 conduction & 0.85 & 1.50 \\
A--AB MoTe$_2$, $5.09^\circ$ & highest 10 valence & 0.29 & 1.01 \\
\bottomrule
\end{tabular}
\end{table*}

\clearpage

\section{Low-energy reduction and symmetry}
\label{app:low_energy_symmetry}

The low-energy projection and symmetry transfer build on Ref.~\cite{zhang_universal_2024}.  This appendix gives their derivation. The purpose is to reduce the large TAPW action to two objects: the channel map \(Q\mapsto Q_g\) and a small matrix acting on the selected layer, orbital, and spin degrees of freedom.  The derivation applies to unitary and antiunitary operations through the notation \(A^{[\eta_g]}=A\) or \(A^*\).

\subsection{Reference-state basis}
\label{app:subsec:reference_orbital_basis}
\label{app:subsec:reference_diagnostics}

The main text uses \(\varsigma\) directly when the fixed-\(\bQ\) blocks of different valley--twist sectors are diagonalized separately and the continuum basis is associated one-to-one with those sectors. For the general construction, let \(\beta=\{\varsigma_1,\ldots,\varsigma_m\}\) denote sectors whose internal states are combined and diagonalized jointly at the same \(\bQ\). For a block containing one sector, \(\beta=\{\varsigma\}\). When several sectors share the same set of \(\bQ\) vectors, as in the \(\Gamma\)-valley examples, the combined block includes all matrix elements among these sectors at fixed \(\bQ\), including their off-diagonal couplings. Let \(n_\beta\) be the dimension of this block and \(r_\beta\) the number of selected states. Its selected eigenvectors define
\begin{equation}
  \begin{aligned}
    h_{\beta,\bQ}E_{\beta,\bQ}^{\rm sel}
    &=E_{\beta,\bQ}^{\rm sel}
      \varepsilon_{\beta,\bQ}^{\rm sel},
    \qquad
    E_{\beta,\bQ}^{{\rm sel}\dagger}
    E_{\beta,\bQ}^{\rm sel}=I_{r_\beta},\\
    p_{\beta,\bQ}(\kvec)
    &=E_{\beta,\bQ}^{\rm sel}(\kvec)
      E_{\beta,\bQ}^{{\rm sel}\dagger}(\kvec).
  \end{aligned}
  \label{app:eq:joint_qblock_projector}
\end{equation}
The full low-energy projector is obtained by embedding these block projectors in the TAPW space and summing over the jointly diagonalized fixed-\(\bQ\) blocks.

For \(y=(\beta,\bQ)\), let \(J_y\) embed its \(n_\beta\) internal coordinates into the full TAPW space.  Appendix~A gives
\begin{equation}
  D_g^{\rm TAPW}J_y^{[\eta_g]}
  =
  J_{\sigma_g(y)}d_{g,\beta},
  \qquad
  \sigma_g(y)=\bigl(g\beta,Q_g(\beta,\bQ)\bigr),
  \label{app:eq:low_source_fiber_factorization}
\end{equation}
where \(d_{g,\beta}\) is unitary and independent of momentum and \(Q\).  The same statement covers a single valley--twist sector or several sectors diagonalized jointly. To make the momentum dependence explicit, set \(x=(\kvec,y)\) and \(x'=(\mathcal R_g\kvec,\sigma_g(y))\).

Let \(p_x\) be the projector onto the \(r_\beta\) selected low-energy states inside this internal space, and let \(a_\beta\) contain \(r_\beta\) mutually orthonormal reference states defined by their layer, orbital, and spin components in the same orthonormalized TAPW coordinates. Thus \(a_\beta^\dagger a_\beta=I_{r_\beta}\). Following the projected-orbital construction of Refs.~\cite{damle_scdm_2015,damle_scdmk_2017}, the reference-orbital frame is
\begin{equation}
  u_x
  =
  p_xa_\beta
  \left(a_\beta^\dagger p_xa_\beta\right)^{-1/2},
  \qquad
  u_x^\dagger u_x=I .
  \label{app:eq:local_scdm_frame}
\end{equation}
It exists when the projected reference orbitals are linearly independent. The full TAPW frames are \(U_x=J_yu_x\) and \(U_{x'}=J_{\sigma_g(y)}u_{x'}\). For fixed reference orbitals \(a_\beta\), Eq.~\eqref{app:eq:local_scdm_frame} depends on the selected states only through \(p_x\), so unitary rotations among degenerate eigenvectors do not change the resulting basis.

The symmetry transfer uses two physical requirements.  First, the selected states at \(x\) are mapped into the selected states at \(x'\) defined above. Second, the ordered reference orbitals at the two ends represent the same symmetry-related low-energy degrees of freedom.  In equations,
\begin{equation}
  p_{x'}
  =
  d_{g,\beta}p_x^{[\eta_g]}d_{g,\beta}^\dagger,
  \qquad
  d_{g,\beta}a_\beta^{[\eta_g]}
  =
  a_{g\beta}\rho_{g,\beta}.
  \label{app:eq:projector_and_reference_covariance}
\end{equation}
The second relation requires the transformed reference states to lie in the column space of \(a_{g\beta}\). Their ordering fixes the internal action \(\rho_{g,\beta}\). Since \(a_\beta\), \(a_{g\beta}\), and \(d_{g,\beta}\) are independent of momentum and \(Q\), so is \(\rho_{g,\beta}\). Moreover, \(a_\beta^\dagger a_\beta=a_{g\beta}^\dagger a_{g\beta}=I\) and the unitarity of \(d_{g,\beta}\) give \(\rho_{g,\beta}^\dagger\rho_{g,\beta}=I\).

To show explicitly how the projected frame transforms, define its positive overlap matrix
\begin{equation}
  c_x=a_\beta^\dagger p_xa_\beta .
  \label{app:eq:anchor_overlap_matrix}
\end{equation}
The two covariance relations in Eq.~\eqref{app:eq:projector_and_reference_covariance} imply
\begin{equation}
  c_{x'}
  =
  \rho_{g,\beta}c_x^{[\eta_g]}\rho_{g,\beta}^\dagger,
  \qquad
  c_{x'}^{-1/2}\rho_{g,\beta}
  =
  \rho_{g,\beta}(c_x^{-1/2})^{[\eta_g]}.
  \label{app:eq:anchor_overlap_covariance}
\end{equation}
The second identity follows from the spectral definition of the inverse square root and the unitary similarity in the first identity.  Substitution into Eq.~\eqref{app:eq:local_scdm_frame} gives
\begin{equation}
  \begin{aligned}
  d_{g,\beta}u_x^{[\eta_g]}
  &=d_{g,\beta}p_x^{[\eta_g]}a_\beta^{[\eta_g]}
    (c_x^{-1/2})^{[\eta_g]}\\
  &=p_{x'}a_{g\beta}\rho_{g,\beta}
    (c_x^{-1/2})^{[\eta_g]}\\
  &=p_{x'}a_{g\beta}c_{x'}^{-1/2}\rho_{g,\beta}
   =u_{x'}\rho_{g,\beta}.
  \end{aligned}
  \label{app:eq:local_frame_covariance}
\end{equation}
Multiplication by \(u_{x'}^\dagger\), followed by lifting through \(J_y\) and \(J_{\sigma_g(y)}\), yields
\begin{equation}
  U_{x'}^\dagger D_g^{\rm TAPW}U_x^{[\eta_g]}
  =
  \rho_{g,\beta}.
  \label{app:eq:q_uniform_route_block}
\end{equation}
This equality holds for every \(x\), so the projected action contains the same \(\rho_{g,\beta}\) at every symmetry-related momentum and channel. After the reference columns are grouped into continuum sectors \(\lambda\), let \(d_\lambda\) be the number of internal states in sector \(\lambda\), and let \(\Lambda_\beta\) be the set of continuum sectors obtained from \(\beta\). Their dimensions satisfy
\begin{equation}
  r_\beta
  =
  \sum_{\lambda\in\Lambda_\beta}d_\lambda.
  \label{app:eq:joint_rank_decomposition}
\end{equation}
When each continuum sector is associated one-to-one with a valley--twist sector, we identify the label \(\lambda\) with \(\varsigma\), set \(d_\lambda=r_\varsigma\) and \(\mathcal Q_\lambda=\mathcal Q_\varsigma\), and recover the notation of the main text. In this case, \(r_\beta=\sum_{\varsigma\in\beta}r_\varsigma\). For reference-state groups permuted as a whole by symmetry, the full projected sewing matrix is
\begin{equation}
  \boxed{
  \left[B_{g,j}^{\rm low}\right]_{
    (\lambda',\xi',\bQ'),(\lambda,\xi,\bQ)}
  =
  \delta_{\lambda',\,\lambda_g(\lambda)}
  \delta_{\bQ',\,Q_g(\lambda,\bQ)}
  [\rho_{g,\lambda}]_{\xi'\xi}.
  }
  \label{app:eq:q_uniform_projected_action}
\end{equation}
Thus \(Q\) labels repeated plane-wave components of the same internal low-energy fields: symmetry permutes those components and applies \(\rho_{g,\lambda}\) to their internal labels.

Reference-orbital selection fixes coordinates inside an already selected low-energy space. Let \(E_x\) contain the selected orthonormal states of a fixed-\(Q\) block, so that \(p_x=E_xE_x^\dagger\). For the fixed ordered reference pattern \(a_\beta\),
\begin{equation}
  \begin{aligned}
    c_x&=a_\beta^\dagger p_xa_\beta
       =(E_x^\dagger a_\beta)^\dagger(E_x^\dagger a_\beta),\\
    \lambda_{\min}(c_x)
       &=\sigma_{\min}(E_x^\dagger a_\beta)^2
        =\min_{\|z\|=1}\|p_xa_\beta z\|^2.
  \end{aligned}
  \label{app:eq:reference_conditioning_metric}
\end{equation}
A small minimum eigenvalue indicates that a linear combination of the references has only a small projection onto the selected subspace. The full reference overlap matrix is block diagonal, so its minimum eigenvalue is the minimum over the fixed-\(Q\) blocks.

MoireKP uses column-pivoted QR at a representative momentum to select seed orbitals. The seeds are retained individually or combined into orbital pairs with sector-dependent phase conventions, using their projected weights in the selected subspace to compare candidate combinations. Once the reference pattern and layer/spin ordering are fixed, the same references are used across momenta and \(Q\) channels. Conditioning is checked over all retained channels and construction momenta, and reference-space covariance is checked separately.

\begin{table*}[t]
\centering
\small
\caption{Reference-state conditioning and symmetry covariance for the six models. \(\lambda_*\) is the minimum of Eq.~\eqref{app:eq:reference_conditioning_metric} over all retained channels and contributing blocks along \(\Gamma-M-K-\Gamma\). \(\epsilon_a^{\max}\) is the maximum residual in Eq.~\eqref{app:eq:raw_reference_residual} over the retained symmetry generators, reference sectors, and channels. Residuals below \(10^{-14}\) are displayed as zero.}
\label{tab:reference_conditioning}
\begin{tabular}{lrr}
\toprule
Model & \(\lambda_*\) & \(\epsilon_a^{\max}\) \\
\midrule
Bilayer MoTe$_2$ \(K\) & 0.280848 & \(2.68\times10^{-6}\) \\
Bilayer ZrS$_2$ \(M_1\) & 0.080051 & \(1.54\times10^{-11}\) \\
Bilayer ZrS$_2$ \(\Gamma\) & 0.001407 & \(0\) \\
Trilayer MoTe$_2$ \(K_A\) & 0.230072 & \(2.40\times10^{-6}\) \\
Trilayer MoTe$_2$ \(K_B\) & 0.237879 & \(2.40\times10^{-6}\) \\
Trilayer MoTe$_2$ \(\Gamma\) & 0.005838 & \(4.39\times10^{-13}\) \\
\bottomrule
\end{tabular}
\end{table*}

For the covariance check, embed the reference states as \(\mathcal A_y=J_y a_\beta\), with \(y=(\beta,Q)\) and \(y'=\sigma_g(y)\). We evaluate
\begin{equation}
  \epsilon^{(a)}_{g,y}
  =\frac{\bigl\|(I-\mathcal A_{y'}\mathcal A_{y'}^\dagger)
                  D_g^{\rm TAPW}\mathcal A_y^{[\eta_g]}\bigr\|_F}
         {\bigl\|D_g^{\rm TAPW}\mathcal A_y^{[\eta_g]}\bigr\|_F}.
  \label{app:eq:raw_reference_residual}
\end{equation}
This residual measures the component of the transformed references outside the corresponding reference space. Table~\ref{tab:reference_conditioning} reports the minimum overlap eigenvalues and maximum covariance residuals for all six models. The projected reference states remain linearly independent on the sampled paths.

\begin{figure*}[t]
  \centering
  \includegraphics[width=0.98\textwidth]{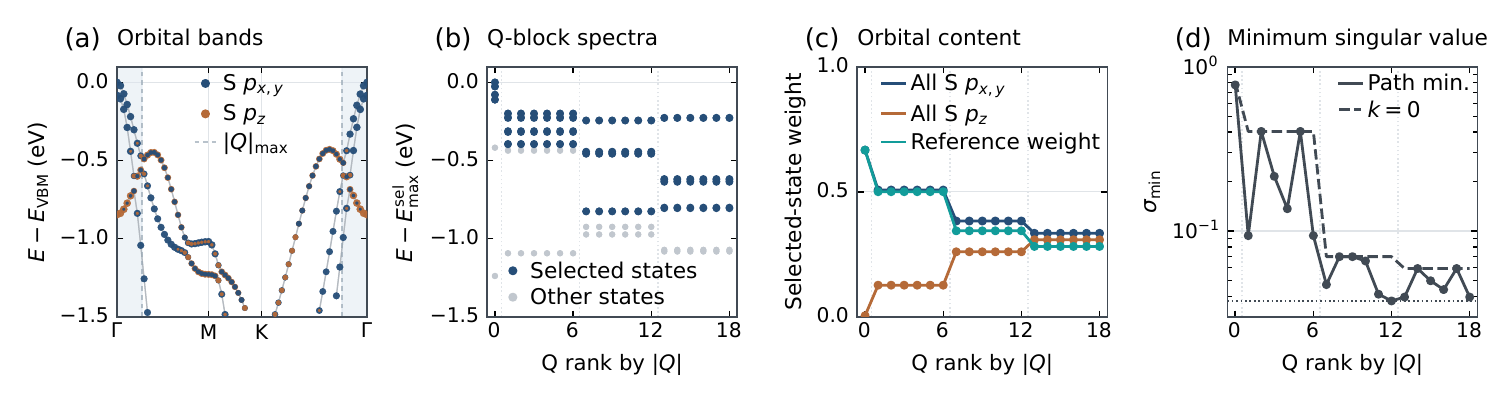}
  \caption{Orbital content and reference-state conditioning for ZrS$_2$ \(\Gamma\). (a) First-radial S \(p_x/p_y\) and \(p_z\) monolayer populations, defined in Eq.~\eqref{app:eq:reference_monolayer_population}, with energies relative to the monolayer VBM; dashed boundaries mark the maximum \(Q\) radius. (b) Fixed-\(Q\) block eigenvalues at \(\kvec=0\), highlighting the eight selected states, with energies relative to the maximum selected energy over the channels. (c) Mean weights at \(\kvec=0\) in the first-radial \(p_x/p_y\), \(p_z\), and symmetric-reference spaces defined in Eq.~\eqref{app:eq:zrs2_orbital_and_reference_weights}. (d) Smallest reference singular value at \(\kvec=0\) (dashed) and its minimum along the high-symmetry \(k\) path (solid). Channels in (b--d) are ordered by increasing \(|Q|\).}
  \label{fig:reference_zrs2_monolayer}
  \label{fig:reference_zrs2_qblocks}
\end{figure*}

\begin{figure*}[t]
 \centering
 \includegraphics[width=0.98\textwidth]{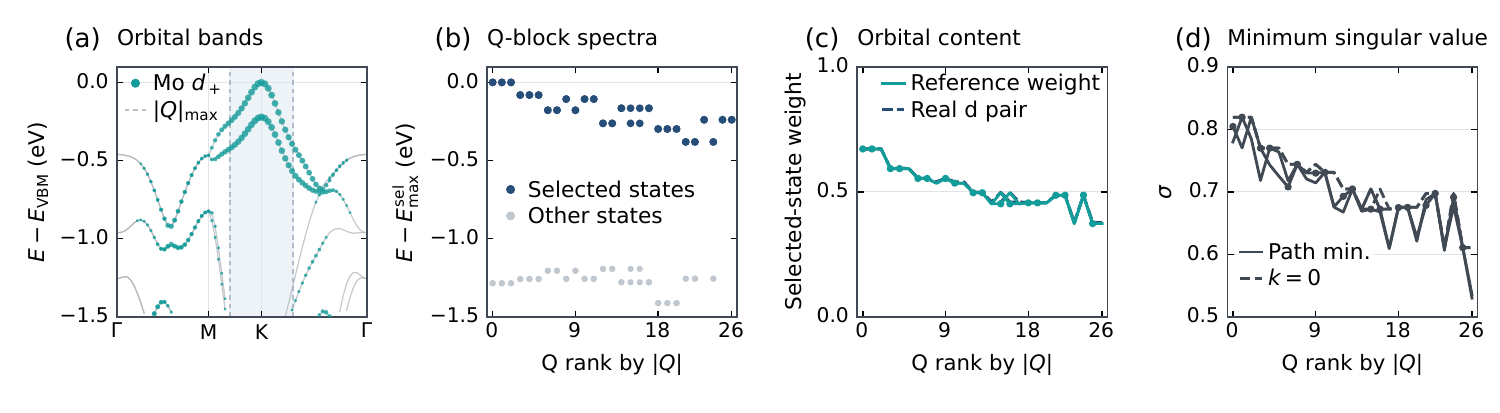}
 \caption{Orbital content and reference-state conditioning for bilayer MoTe$_2$ \(K\). (a) Monolayer bands with Mo \(d_+\) coefficient weights, using the convention of Fig.~\ref{fig:mote2_k_flagship}(a) and the monolayer VBM zero; dashed boundaries mark the maximum \(Q\) radius. (b) Spin-up layer-block eigenvalues at \(\kvec=0\), highlighting the selected state in each block, with energies relative to the maximum selected energy over both layers and all channels. (c) \(d_+\) reference weight and total \(d_{x^2-y^2}/d_{xy}\) weight at \(\kvec=0\). (d) Reference singular value at \(\kvec=0\) (dashed) and its minimum along the high-symmetry \(k\) path (solid). Both layers are shown, with channels ordered independently by increasing \(|Q|\) in (b--d).}
 \label{fig:reference_mote2_good}
\end{figure*}

Figures~\ref{fig:reference_zrs2_qblocks} and \ref{fig:reference_mote2_good} illustrate how the orbital content of the selected states affects the reference overlaps. For the ZrS$_2$ \(\Gamma\) model, eight states are selected jointly from each spinful two-layer fixed-\(Q\) block. At the reference momentum they form four Kramers pairs predominantly of S \(p_x/p_y\) character. The reference states are symmetric combinations of the two S sites in each layer,

\begin{equation}
  |a_{\ell s m}^{(+)}\rangle
  =\frac{|S_{1\ell},p_m,r_1,s\rangle
              +|S_{2\ell},p_m,r_1,s\rangle}{\sqrt{2}},
  \qquad m=x,y.
  \label{app:eq:zrs2_reference_orbitals}
\end{equation}
Here \(\ell=1,2\), \(s=\uparrow,\downarrow\), and \(r_1\) denotes the first radial basis function. The eight references are ordered by layer \((1,2)\), then spin \((\downarrow,\uparrow)\), and then orbital \((p_x,p_y)\).

For the monolayer orbital populations in Fig.~\ref{fig:reference_zrs2_monolayer}(a), a generalized eigenvector \(z_n\) gives
\begin{equation}
  w_{\mu n}^{\rm loc}
  =\frac{\max\{0,\operatorname{Re}[z_{\mu n}^*(S^{\rm loc}z_n)_\mu]\}}
         {\sum_\nu\max\{0,\operatorname{Re}[z_{\nu n}^*(S^{\rm loc}z_n)_\nu]\}}.
  \label{app:eq:reference_monolayer_population}
\end{equation}
In the orthonormal TAPW coordinates, let \(\Pi_{xy}\) and \(\Pi_z\) project onto the first-radial S \(p_x/p_y\) and \(p_z\) components of both layers and spins, and let \(\Pi_a=a_\beta a_\beta^\dagger\). The mean weights over the eight selected states in Fig.~\ref{fig:reference_zrs2_qblocks}(c) are
\begin{equation}
  \begin{aligned}
    W_{xy}(x)&=\tfrac18\operatorname{Tr}(p_x\Pi_{xy}),\qquad
    W_z(x)=\tfrac18\operatorname{Tr}(p_x\Pi_z),\\
    W_a(x)&=\tfrac18\operatorname{Tr}(p_x\Pi_a)
           =\tfrac18\operatorname{Tr}(c_x).
  \end{aligned}
  \label{app:eq:zrs2_orbital_and_reference_weights}
\end{equation}
Since the reference space contains only symmetric S-site combinations, \(W_{xy}-W_a\) is the weight in the corresponding antisymmetric combinations. In outer \(Q\) channels, the selected states acquire more \(p_z\) character and smaller \(p_x/p_y\) and reference weights [Fig.~\ref{fig:reference_zrs2_qblocks}(b,c)], accompanied by smaller minimum reference overlaps [Fig.~\ref{fig:reference_zrs2_qblocks}(d)]. Projected reference vectors can also become nearly linearly dependent despite retaining finite individual weights, so the mean reference weight alone does not characterize their conditioning.

For the bilayer MoTe$_2$ \(K\) model, one state is selected from each spin-up layer block. The reference on each layer is the spin-up Mo \(d_+\) orbital with the first radial basis function, where \(d_+=(d_{x^2-y^2}+\mathrm{i}d_{xy})/\sqrt2\).

For this rank-one selection, \(c_x\) is the squared overlap between the reference and the selected local state. The reference weight nearly coincides with the total \(d_{x^2-y^2}/d_{xy}\) weight [Fig.~\ref{fig:reference_mote2_good}(c)], and the reference overlap remains well separated from zero along the sampled path [Fig.~\ref{fig:reference_mote2_good}(d)].

\subsection{Energy-linearized reduction}
\label{app:subsec:downfold_sensitivity}

We omit the reference-basis superscript and momentum argument in this subsection. Eliminating the excluded-space component from the eigenproblem of \(H=\Horth\) gives
\begin{equation}
 \begin{aligned}
 \bigl[H_{LL}+\Sigma(E)\bigr]c_L&=E c_L,\\
 \Sigma(E)&=H_{LH}(E I-H_{HH})^{-1}H_{HL}.
 \end{aligned}
 \label{app:eq:energy_dependent_schur}
\end{equation}
Let \(E_0=E_{\rm ref}\), \(R_0=(E_0 I-H_{HH})^{-1}\), \(\Sigma_0=H_{LH}R_0H_{HL}\), and \(D=H_{LH}R_0^2H_{HL}\). Expanding about \(E_0\) gives
\begin{equation}
 \Sigma(E)=\Sigma_0-(E-E_0)D+O((E-E_0)^2).
 \label{app:eq:schur_linearization}
\end{equation}
Keeping this first-order term gives the generalized eigenproblem
\begin{equation}
 \begin{aligned}
 A_0c_L&=E S_{\rm eff}c_L,\\
 A_0&=H_{LL}+\Sigma_0+E_0D,\\
 S_{\rm eff}&=I+D.
 \end{aligned}
 \label{app:eq:linearized_generalized_eigenproblem}
\end{equation}
For Hermitian \(H_{HH}\) and real \(E_0\) outside its spectrum, \(D\) is positive semidefinite and \(S_{\rm eff}\) is positive definite. With \(v=S_{\rm eff}^{1/2}c_L\), the final static Hermitian Hamiltonian is
\begin{equation}
 H_{\rm eff}^{\rm lin}=S_{\rm eff}^{-1/2}A_0S_{\rm eff}^{-1/2},
 \label{app:eq:linearized_static_hamiltonian}
\end{equation}
which is Eq.~\eqref{eq:projected_heff_linearized}, used for all six models. The fixed-energy form in Eq.~\eqref{eq:projected_heff} instead retains only \(H_{LL}+\Sigma_0\).

The corresponding normalized reconstruction within the chosen input space is
\begin{equation}
 \begin{aligned}
 \Psi_{\rm lin}&=(U_L+U_H X_0)S_{\rm eff}^{-1/2}v,\\
 X_0&=R_0H_{HL},\qquad \|v\|=1.
 \end{aligned}
 \label{app:eq:linearized_state_reconstruction}
\end{equation}
Writing \(W_0=(U_L+U_HX_0)(I+X_0^\dagger X_0)^{-1/2}\), the block relation \(H_{HH}X_0=E_0X_0-H_{HL}\) gives
\begin{equation}
 W_0^\dagger W_0=I,\qquad H_{\rm eff}^{\rm lin}=W_0^\dagger H W_0.
 \label{app:eq:normalized_reconstruction_projection}
\end{equation}
Thus \(H_{\rm eff}^{\rm lin}\) is a Rayleigh--Ritz projection onto the reconstructed space evaluated at \(E_0\). For \(K_A/K_B\), reconstruction is confined to the retained spin sector. The overlaps in Figs.~\ref{fig:mote2_k_flagship}--\ref{fig:mote2_aab_multivalley} instead use \(U_Lv^{\rm cont}\), without this reconstruction or renormalization by projected TAPW weight (Sec.~\ref{sec:validation_protocol}).

The reference energies and target-band sets are listed in Table~\ref{tab:paper_model_configurations}. For a fixed target set \(\mathcal I\), the reference-energy scan compares \(H_{\rm eff}\) with TAPW through
\begin{equation}
 \begin{aligned}
 \epsilon_{\rm RMS}(E_0)^2
 &=\frac{1}{N_k|\mathcal I|}\sum_{j=1}^{N_k}\sum_{n\in\mathcal I}\bigl[\delta\epsilon_{nj}(E_0)\bigr]^2,\\
 \delta\epsilon_{nj}(E_0)
 &=\epsilon_n^{\rm eff}(\kvec_j;E_0)
   -\epsilon_{m(n,j)}^{\rm TAPW}(\kvec_j).
 \end{aligned}
 \label{app:eq:downfold_scan_rms}
\end{equation}
Samples have equal weight, and the band correspondence is held fixed during the scan. The raw RMS uses unshifted energies. The aligned RMS subtracts the constant band-edge difference \(s(E_0)=E_{\rm edge}^{\rm eff}(E_0)-E_{\rm edge}^{\rm TAPW}\), where the edge is the maximum valence or minimum conduction energy over the target bands and comparison path:
\begin{equation}
 \epsilon_{\rm aligned}(E_0)^2
 =\frac{\sum_{j,n}[\delta\epsilon_{nj}(E_0)-s(E_0)]^2}
        {N_k|\mathcal I|}.
 \label{app:eq:downfold_aligned_rms}
\end{equation}
The \(K_A/K_B\) sectors share the \(K_A\) edge shift, preserving their relative offset, and use the joint full-spin TAPW state assignment defined in Sec.~\ref{sec:validation_protocol}. Their comparison with full-spin TAPW includes the spin-block approximation as well as low-energy reduction.

We also measure separation from the eliminated spectrum. At each sampled momentum let \(\mathcal W_j\) be the interval spanned by the selected TAPW energies. Define
\begin{equation}
 \Delta_{\rm excl}=\min_j\operatorname{dist}
       \bigl(\mathcal W_j,\operatorname{spec}H_{HH}(\kvec_j)\bigr).
 \label{app:eq:excluded_window_distance}
\end{equation}
Table~\ref{tab:downfold_reference_scan} reports the target energy intervals and their minimum separation from the excluded-block spectrum.

\begin{figure*}[!t]
 \centering
 \includegraphics[width=0.98\textwidth]{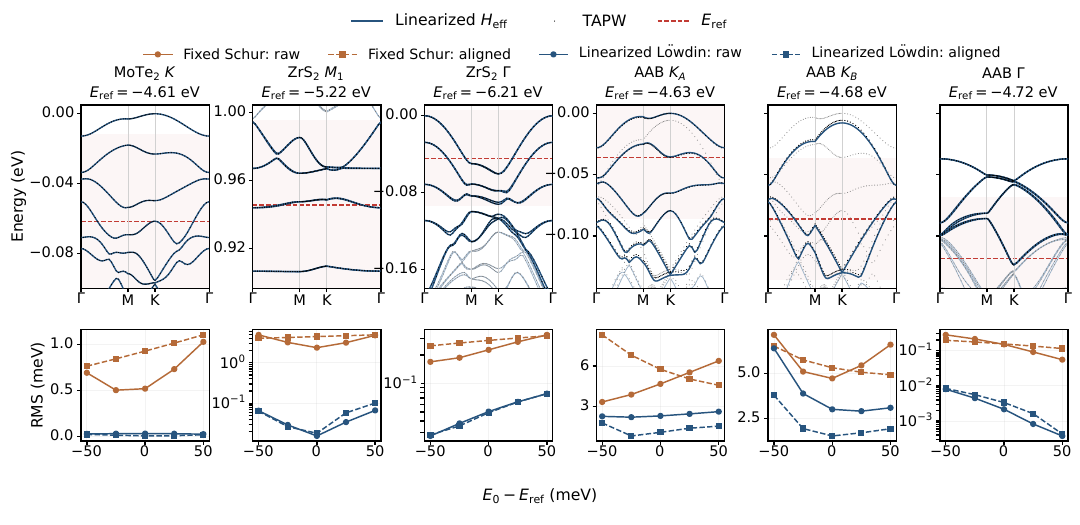}
 \caption{Band comparison and reference-energy sensitivity for the six valley models along $\Gamma-M-K-\Gamma$. Upper panels compare TAPW (dots) and energy-linearized $H_{\rm eff}$ (lines); black and blue highlight the target bands, while gray and pale blue show the remaining bands. Energy zeros are the TAPW VBM for bilayer MoTe$_2$, the TAPW $\Gamma$ VBM for ZrS$_2$, and the TAPW $K_A$ VBM for trilayer MoTe$_2$. The bilayer MoTe$_2$ $H_{\rm eff}$ VBM is aligned to TAPW; $K_A/K_B$ share a $K_A$-derived shift, while ZrS$_2$ and trilayer $\Gamma$ receive no extra shift. Red dashed lines mark $E_{\rm ref}$ and shading marks the scan interval. Lower panels show raw (solid) and band-edge-aligned (dashed) RMS differences for the fixed-energy and energy-linearized reductions in the same target windows.}
 \label{fig:downfold_reference_sensitivity}
 \label{fig:downfold_bands_eref}
\end{figure*}

Figure~\ref{fig:downfold_reference_sensitivity} compares the fixed-energy and energy-linearized reductions while varying the reference energy by up to \(\pm50\) meV. The selected subspaces and reference states remain fixed, and no continuum coefficients are re-extracted. The \(K_A/K_B\) reference energies receive the same offset, with a common \(K_A\) alignment for each pair. At the chosen reference energies, the energy-linearized form gives smaller RMS differences in the target windows for all six models.

\begin{table*}[t]
\centering
\small
\setlength{\tabcolsep}{6pt}
\caption{Target energy intervals and excluded-state separations for the band sets in Table~\ref{tab:paper_model_configurations}. The intervals span the selected TAPW energies along the high-symmetry path; $\Delta_{\rm excl}$ is defined in Eq.~\eqref{app:eq:excluded_window_distance}.}
\label{tab:downfold_reference_scan}
\begin{tabular}{lcr}
\toprule
Model & TAPW interval (eV) & $\Delta_{\rm excl}$ (eV) \\
\midrule
Bilayer MoTe$_2$ $K$ & $[-4.68,-4.55]$ & 0.94 \\
ZrS$_2$ $M_1$ & $[-5.26,-5.17]$ & 1.52 \\
ZrS$_2$ $\Gamma$ & $[-6.35,-6.17]$ & 0.23 \\
Trilayer MoTe$_2$ $K_A$ & $[-4.73,-4.60]$ & 0.12 \\
Trilayer MoTe$_2$ $K_B$ & $[-4.73,-4.60]$ & 0.11 \\
Trilayer MoTe$_2$ $\Gamma$ & $[-4.72,-4.64]$ & 0.52 \\
\bottomrule
\end{tabular}
\end{table*}

\subsection{Symmetry representations}
\label{app:subsec:group_relations}
\label{app:subsec:continuum_sewing}

The projected matrices inherit the multiplication law of the TAPW symmetry action.  If
\begin{equation}
  d_{g,h\beta}d_{h,\beta}^{[\eta_g]}
  =
  \omega(g,h)d_{gh,\beta},
  \qquad |\omega(g,h)|=1,
  \label{app:eq:source_internal_group_relation}
\end{equation}
then applying Eq.~\eqref{app:eq:projector_and_reference_covariance} twice to the same reference pattern gives
\begin{equation}
  \begin{aligned}
  d_{g,h\beta}d_{h,\beta}^{[\eta_g]}
  a_\beta^{[\eta_{gh}]}
  &=a_{gh\beta}
    \rho_{g,h\beta}\rho_{h,\beta}^{[\eta_g]}\\
  &=\omega(g,h)a_{gh\beta}\rho_{gh,\beta},
  \qquad
  \eta_{gh}=\eta_g+\eta_h\pmod 2.
  \end{aligned}
  \label{app:eq:composed_reference_action}
\end{equation}
Since \(a_{gh\beta}\) has full column rank, the two internal factors must agree:
\begin{equation}
  \rho_{g,h\beta}\rho_{h,\beta}^{[\eta_g]}
  =
  \omega(g,h)\rho_{gh,\beta}.
  \label{app:eq:anchor_group_relation}
\end{equation}

The same projected action also constrains the effective Hamiltonian.  Let \(U_{L,r}\) and \(U_{H,r}\) be orthonormal frames for the selected and eliminated subspaces at \(r={\rm s},{\rm t}\), and write
\begin{equation}
  D_g^{\rm TAPW}U_{X,{\rm s}}^{[\eta_g]}
  =U_{X,{\rm t}}\mathcal U_{g,X},
  \qquad X=L,H,
  \qquad \mathcal U_{g,L}=B_{g,j}^{\rm low}.
  \label{app:eq:low_high_frame_covariance}
\end{equation}
Covariance of the selected projector implies covariance of its complement.  Consequently the blocks \(H_{XY}=U_X^\dagger\Horth U_Y\) transform as
\begin{equation}
  H_{XY,{\rm t}}
  =\mathcal U_{g,X}
  H_{XY,{\rm s}}^{[\eta_g]}
  \mathcal U_{g,Y}^\dagger .
  \label{app:eq:schur_block_covariance}
\end{equation}
For the same real \(E_{\rm ref}\) at the two endpoints, define \(R_r=(E_{\rm ref}I-H_{HH,r})^{-1}\).  Functional calculus gives
\begin{equation}
  R_{\rm t}
  =\mathcal U_{g,H}R_{\rm s}^{[\eta_g]}
  \mathcal U_{g,H}^\dagger .
  \label{app:eq:schur_resolvent_covariance}
\end{equation}
Substituting these relations into the fixed-energy downfolding shows the cancellation of the eliminated-subspace action explicitly:
\begin{equation}
  \begin{aligned}
  H_{{\rm eff},{\rm t}}^{\rm ref}
  &=H_{LL,{\rm t}}+H_{LH,{\rm t}}R_{\rm t}H_{HL,{\rm t}}\\
  &=B_{g,j}^{\rm low}
    [H_{LL,{\rm s}}+H_{LH,{\rm s}}R_{\rm s}H_{HL,{\rm s}}]^{[\eta_g]}
    B_{g,j}^{{\rm low}\dagger}\\
  &=B_{g,j}^{\rm low}
    [H_{{\rm eff},{\rm s}}^{\rm ref}]^{[\eta_g]}
    B_{g,j}^{{\rm low}\dagger}.
  \end{aligned}
  \label{app:eq:effective_hamiltonian_covariance}
\end{equation}
The energy-linearized form in Eq.~\eqref{eq:projected_heff_linearized} has the same covariance. Since \(D_{\rm t}=B_{g,j}^{\rm low}D_{\rm s}^{[\eta_g]}B_{g,j}^{{\rm low}\dagger}\), the effective overlap obeys \(S_{{\rm eff},{\rm t}}=B_{g,j}^{\rm low}S_{{\rm eff},{\rm s}}^{[\eta_g]}B_{g,j}^{{\rm low}\dagger}\). Its positive inverse square root transforms in the same way. Thus covariance requires a covariant TAPW Hamiltonian and selected projector, a common real reference energy, and an existing high-energy resolvent.

The reference-state basis constructed above is used for both \(H_{\rm eff}\) and the projected symmetry matrices. Any fixed rotation used to order the internal states is absorbed into this convention and repeated in every \(Q\) channel. Thus Eq.~\eqref{app:eq:q_uniform_projected_action} directly gives the internal matrix \(\rho_{g,\lambda}\).

For each generator, one selected symmetry-related momentum pair supplies \(\rho_{g,\lambda}\). Additional momentum pairs, when present, test projected-subspace leakage, projected-action unitarity, and Eq.~\eqref{app:eq:effective_hamiltonian_covariance} at other momenta. The valley-sector map, allowed matrix support, and \(Q\)-channel map are fixed before the numerical correction.

With \(P_{L,{\rm s}}\) and \(P_{L,{\rm t}}\) denoting the full selected projectors, the subspace leakage and projected-action unitarity are measured by
\begin{equation}
  \begin{aligned}
  \epsilon_{g}^{\rm leak}
  &=
  \frac{\|(I-P_{L,{\rm t}})D_g^{\rm TAPW}
  P_{L,{\rm s}}^{[\eta_g]}\|_{\rm F}}
  {\|P_{L,{\rm s}}\|_{\rm F}},\\
  \epsilon_g^{\rm U,proj}
  &=
  \max_j
  \|B_{g,j}^{{\rm low}\dagger}B_{g,j}^{\rm low}-I\|_{\rm F}.
  \end{aligned}
  \label{app:eq:projection_diagnostics}
\end{equation}
These quantities are evaluated at the symmetry-related momentum pairs before correction.

We make a small adjustment within the specified internal blocks so that unitarity, generator powers and the semilinear group relations hold within their numerical tolerances. The correction acts within the selected subspace and leaves any component outside it unchanged.

The semilinear multiplication used by the correction is
\begin{equation}
  (\rho_g\mathcal K^{\eta_g})
  (\rho_h\mathcal K^{\eta_h})
  =
  \rho_g\rho_h^{[\eta_g]}
  \mathcal K^{\eta_g+\eta_h}.
  \label{app:eq:semilinear_composition}
\end{equation}
Generator powers and group relations are evaluated with the same conjugation rule as Eq.~\eqref{app:eq:anchor_group_relation}.  If \(\rho\) and \(\rho^{\rm cont}\) denote the internal blocks before and after correction, the internal-block residuals and correction are
\begin{equation}
  \begin{aligned}
  \delta_{\rm corr}
  &=\max_{g,\lambda}
  \frac{\|\rho_{g,\lambda}^{\rm cont}
  -\rho_{g,\lambda}\|_{\rm F}}
  {\max(1,\|\rho_{g,\lambda}\|_{\rm F})},\\
  \epsilon_{\rm U}
  &=\max_{g,\lambda}
  \|\rho_{g,\lambda}^{{\rm cont}\dagger}
  \rho_{g,\lambda}^{\rm cont}-I\|_{\rm F},\\
  \epsilon_{\rm grp}
  &=\max_R
  \|\rho^{\rm cont}(w_{R,L})
  -\omega_R\rho^{\rm cont}(w_{R,R})\|_{\rm F}.
  \end{aligned}
  \label{app:eq:correction_diagnostics}
\end{equation}
Here \(\rho(w_{R,L})=\omega_R\rho(w_{R,R})\), with each word evaluated along its sector sequence and \(Q\)-channel maps. For the full matrices, the correction and unitarity residuals are \(\|B_g^{\rm low}-U_g^{\rm cont}\|_F/\sqrt{N_{\rm cont}}\) and \(\|A^\dagger A-I\|_F/\sqrt{N_{\rm cont}}\), respectively, where \(A\) denotes the projected or corrected action.

For unitary \(D_g\) and orthonormal source and target frames, set \(B=U_t^\dagger D_gU_s^{[\eta_g]}\) and \(L=(I-U_tU_t^\dagger)D_gU_s^{[\eta_g]}\), and let \(\widetilde B\) denote the corrected low-space action. Orthogonal decomposition gives
\begin{equation}
 \begin{aligned}
 I-B^\dagger B&=L^\dagger L,\\
 \|D_gU_s^{[\eta_g]}-U_t\widetilde B\|_F^2
 &=\|L\|_F^2+\|B-\widetilde B\|_F^2.
 \end{aligned}
 \label{app:eq:leakage_correction_decomposition}
\end{equation}

The corrected internal matrices and the unchanged \(Q\)-channel map define
\begin{equation}
  \boxed{
  [U_g^{\rm cont}]_{
    (\lambda',\xi',\bQ'),(\lambda,\xi,\bQ)}
  =
  \delta_{\lambda',\,\lambda_g(\lambda)}
  \delta_{\bQ',\,Q_g(\lambda,\bQ)}
  [\rho_{g,\lambda}^{\rm cont}]_{\xi'\xi}.
  }
  \label{app:eq:exact_continuum_action}
\end{equation}

For the six models, effective-Hamiltonian covariance was rechecked after correction using
\[
 \Delta_g(\kvec)=H_{\rm eff}(\kvec)
 -U_g^{\rm cont}\bigl[H_{\rm eff}(\kappa_g(\kvec))\bigr]^{[\eta_g]}
 U_g^{{\rm cont}\dagger}.
\]
Across the six models, the normalized unitarity residual before correction is below \(3.9\times10^{-6}\); after correction, the unitarity, group-relation and applicable antiunitary-square residuals are below \(10^{-14}\). Subspace leakage is below \(7.1\times10^{-5}\), and correction and intertwining residuals are below \(4.8\times10^{-4}\), within the common acceptance tolerance \(\tau_{\rm sym}=10^{-2}\) for the projected symmetry representation. After correction, the relative effective-Hamiltonian covariance residual \(\|\Delta_g\|_F/\|H_{\rm eff}\|_F\) is below \(1.6\times10^{-6}\), and its absolute value \(\|\Delta_g\|_F\) is below 0.07 meV.

The allowed-operator construction in Appendix~C imposes
\begin{equation}
  U_g^{\rm cont}
  \left[
    H_{\rm cont}\!\left(\kappa_g(\kvec)\right)
  \right]^{[\eta_g]}
  U_g^{{\rm cont}\dagger}
  =
  H_{\rm cont}(\kvec).
  \label{app:eq:exact_continuum_covariance}
\end{equation}

\noindent\textit{Reciprocal-boundary sewing and little-group representations.}\enspace The reciprocal-boundary sewing of the continuum model acts on the \(Q\)-channel labels. For a reciprocal-lattice shift \(\mathbf b\), define
\begin{equation}
  \mathcal S_{\mathbf b}^{\rm mod}
  |\lambda,\xi,\bQ\rangle
  =
  \begin{cases}
    |\lambda,\xi,\bQ+\mathbf b\rangle,
      &\bQ+\mathbf b\in\mathcal Q_\lambda,\\
    0,&\text{otherwise}.
  \end{cases}
  \label{app:eq:model_boundary_sewing}
\end{equation}
Thus reciprocal-boundary sewing shifts the plane-wave channel and leaves the internal index \(\xi\) unchanged. A little-group representation on this space requires the combined symmetry map and reciprocal shift to close on the selected \(Q\) set.

For orthonormal target-band frames \(V_{\rm start}\) and \(V_{\rm end}\), let \(T\) be the corresponding TAPW or continuum map from the endpoint to the starting reciprocal representative. Before polar unitarization, we evaluate
\begin{equation}
 \begin{aligned}
 B&=V_{\rm start}^{\dagger}T V_{\rm end},\\
 \delta_\sigma&=\max_{\kappa_1,a}|\sigma_a(B)-1|,\\
 \ell_{\rm cut}(\kappa_1)&=1-\frac{\|T V_{\rm end}\|_F^2}{r},
 \end{aligned}
 \label{app:eq:boundary_band_diagnostics}
\end{equation}
where \(r=2\) for the displayed highest two bands. The quantity \(\ell_{\rm cut}\) measures their mean norm loss under sewing, while \(\delta_\sigma\) also reflects mismatch with the starting target space. On the TAPW \(81\times81\) mesh, the maximum \(\ell_{\rm cut}\) values for \(n_g=3,4,5\) are \(6.8\times10^{-3}\), \(3.5\times10^{-4}\) and \(1.1\times10^{-5}\), respectively; the corresponding \(\delta_\sigma\) values are \(6.0\times10^{-3}\), \(2.2\times10^{-4}\) and \(9.3\times10^{-6}\). The selected continuum model at its fixed cutoff has \(\delta_\sigma=2.8\times10^{-4}\).

For the TAPW top-two \(\Gamma\)-valley bands, we tested $n_g=3,4,5$ on meshes from $11\times11$ to $81\times81$. For the displayed $n_g=4$ result, refining the mesh from $41\times41$ to $81\times81$ changes the Wilson phases by at most $2.1\times10^{-4}$ cycles; increasing $n_g$ from 4 to 5 at $81\times81$ changes them by at most $4.3\times10^{-4}$ cycles.

For the selected continuum model, we compared 41, 81 and 121 points along the loop at fixed \(Q\) cutoff and 41 transverse points. The maximum 41--121-point phase difference is $2.3\times10^{-4}$ cycles.

If the image of a high-symmetry point under the geometric action differs from the chosen representative by a reciprocal-lattice vector, \(\mathcal R_g\kvec=\kvec+\Delta\bG\), then the little-group action at the chosen representative is
\begin{equation}
  O_g^{\rm KP}(\kvec)
  =
  \mathcal S_{-\Delta\bG}^{\rm mod}U_g^{\rm cont}.
  \label{app:eq:little_group_foldback}
\end{equation}
\(U_g^{\rm cont}\) is independent of \(\kvec\). The momentum dependence here comes from \(\Delta\bG\) and the projected band frame. For a chosen orthonormal band frame \(V_{\mathcal B}(\kvec)\), the band-space representation is
\begin{equation}
  \mathcal D_{g,\mathcal B}^{\rm KP}(\kvec)
  =
  V_{\mathcal B}^\dagger(\kvec)
  O_g^{\rm KP}(\kvec)
  V_{\mathcal B}^{[\eta_g]}(\kvec).
  \label{app:eq:band_space_representation}
\end{equation}
\section{Symmetry-allowed operators}
\label{app:symmetry_allowed_terms}

This appendix constructs a real basis of independent symmetry-allowed Hamiltonian operators for a specified truncation. The plane-wave representation follows Ref.~\cite{zhang_universal_2024}; MoireKP completes the candidate space under symmetry and Hermitian conjugation and extracts an independent invariant basis.

\subsection{Elementary plane-wave terms}
\label{app:subsec:matrix_polynomial_space}
\label{app:real_to_finite_q_terms}

The elementary term used in this appendix is the elementwise matrix \(Y_{ab,\mathbf p}^{mn}\) defined in Eq.~\eqref{eq:raw_elementary_continuum_term}. Its labels specify the internal row and column states, the momentum transfer, and the polynomial order; the Kronecker delta in that equation fixes the matrix support explicitly.

We now derive that form from a local real-space bilinear.  For a Cartesian multi-index \(\bm m=(m_x,m_y)\), write \(|\bm m|=m_x+m_y\), \(\mathbf u^{\bm m}=u_x^{m_x}u_y^{m_y}\), and \(\partial^{\bm m}=\partial_x^{m_x}\partial_y^{m_y}\).  One Fourier component of a general two-sided derivative term is
\begin{equation}
  \widehat Y_{ab,\mathbf p}^{\bm m_{\rm L},\bm m_{\rm R}}
  =
  \int d^2r\,
  \left[i^{|\bm m_{\rm L}|}
  \partial^{\bm m_{\rm L}}\psi_a^\dagger(\mathbf r)\right]
  e^{-i\mathbf p\cdot\mathbf r}
  \left[(-i)^{|\bm m_{\rm R}|}
  \partial^{\bm m_{\rm R}}\psi_b(\mathbf r)\right].
  \label{app:eq:local_bilinear_term}
\end{equation}
For a sample of area \(A\), the plane-wave fields in the selected sectors are
\begin{equation}
  \begin{aligned}
  \psi_b(\mathbf r)
  &=\frac{1}{\sqrt{A}}
  \sum_{\kvec',\bQ_2}
  e^{i(\kvec'-\bQ_2)\cdot\mathbf r}
  \psi_{\kvec',b,\bQ_2},\\
  \psi_a^\dagger(\mathbf r)
  &=\frac{1}{\sqrt{A}}
  \sum_{\kvec,\bQ_1}
  e^{-i(\kvec-\bQ_1)\cdot\mathbf r}
  \psi_{\kvec,a,\bQ_1}^\dagger .
  \end{aligned}
  \label{app:eq:continuum_field_expansion}
\end{equation}
The derivatives supply powers of \(\mathbf u_1=\kvec-\bQ_1\) and \(\mathbf u_2=\kvec-\bQ_2\).  The spatial integral gives
\begin{equation}
  \int_A d^2r\,
  e^{-i(\kvec-\bQ_1)\cdot\mathbf r}
  e^{-i\mathbf p\cdot\mathbf r}
  e^{i(\kvec'-\bQ_2)\cdot\mathbf r}
  =
  A\,\delta_{\kvec',\kvec}
  \delta_{\bQ_1,\bQ_2+\mathbf p}.
  \label{app:eq:real_space_route_orthogonality}
\end{equation}
The corresponding plane-wave matrix is
\begin{equation}
  [Y_{ab,\mathbf p}^{\bm m_{\rm L},\bm m_{\rm R}}(\kvec)]_
  {(a_1,\bQ_1),(a_2,\bQ_2)}
  =
  \delta_{a_1,a}\delta_{a_2,b}
  \delta_{\bQ_1,\bQ_2+\mathbf p}
  (\kvec-\bQ_1)^{\bm m_{\rm L}}
  (\kvec-\bQ_2)^{\bm m_{\rm R}}.
  \label{app:eq:two_sided_elementary_continuum_term}
\end{equation}
This immediately gives
\begin{equation}
  [Y_{ab,\mathbf p}^{\bm m_{\rm L},\bm m_{\rm R}}]^\dagger
  =
  Y_{ba,-\mathbf p}^{\bm m_{\rm R},\bm m_{\rm L}}.
  \label{app:eq:two_sided_elementary_adjoint}
\end{equation}

On the nonzero support, \(\bQ_2=\bQ_1-\mathbf p\), and hence \(\mathbf u_2=\mathbf u_1+\mathbf p\).  The multi-index binomial theorem gives
\begin{equation}
  (\mathbf u+\mathbf p)^{\bm m_{\rm R}}
  =
  \sum_{\bm r\leq\bm m_{\rm R}}
  \binom{\bm m_{\rm R}}{\bm r}
  \mathbf p^{\bm m_{\rm R}-\bm r}\mathbf u^{\bm r},
  \label{app:eq:right_derivative_binomial}
\end{equation}
where the inequality and binomial coefficient act componentwise.  Thus
\begin{equation}
  Y_{ab,\mathbf p}^{\bm m_{\rm L},\bm m_{\rm R}}
  =
  \sum_{\bm r\leq\bm m_{\rm R}}
  \binom{\bm m_{\rm R}}{\bm r}
  \mathbf p^{\bm m_{\rm R}-\bm r}
  Y_{ab,\mathbf p}^{\bm m_{\rm L}+\bm r,\mathbf 0}.
  \label{app:eq:two_sided_to_one_sided}
\end{equation}
The highest order on the right is no larger than \(|\bm m_{\rm L}|+|\bm m_{\rm R}|\).  Conversely, every one-sided term is already a two-sided term with \(\bm m_{\rm R}=0\).  At any fixed maximum order \(M\), the two parametrizations therefore span exactly the same space:
\begin{equation}
  \operatorname{span}_{\mathbb C}
  \{Y_{ab,\mathbf p}^{\bm m_{\rm L},\bm m_{\rm R}}:
  |\bm m_{\rm L}|+|\bm m_{\rm R}|\leq M\}
  =
  \operatorname{span}_{\mathbb C}
  \{Y_{ab,\mathbf p}^{\bm m,\mathbf0}:|\bm m|\leq M\}.
  \label{app:eq:two_sided_one_sided_span}
\end{equation}
Finally, the invertible change of variables \(u_\pm=u_x\pm iu_y\) gives
\begin{equation}
  \operatorname{span}_{\mathbb C}
  \{u_x^{r_x}u_y^{r_y}:r_x+r_y\leq M\}
  =
  \operatorname{span}_{\mathbb C}
  \{u_+^m u_-^n:m+n\leq M\}.
  \label{app:eq:cartesian_circular_polynomial_span}
\end{equation}
With \(u_+=z_{\bQ_1}\) and \(u_-=\bar z_{\bQ_1}\), this proves that the one-sided circular seed in Eq.~\eqref{eq:raw_elementary_continuum_term} loses no term allowed by a general two-sided local derivative expansion at the same total order.

A seed specification \(\mathfrak M\) lists the selected \(Q\) sets, internal matrix connections, harmonic representatives, and maximum polynomial order. MoireKP enumerates Eq.~\eqref{eq:raw_elementary_continuum_term} on this finite seed support; the symmetry-completed support may contain additional partners.  In the one-sided form, Hermitian conjugation exchanges the row and column states and reverses the momentum transfer. The polynomial is re-expanded about the new row channel using
\[
  z_{\bQ+\mathbf p}=z_{\bQ}-(p_x+ip_y),
  \qquad
  \bar z_{\bQ+\mathbf p}=\bar z_{\bQ}-(p_x-ip_y).
\]
This is the circular form of Eq.~\eqref{app:eq:right_derivative_binomial}; it requires only monomials up to the original total order, provided all lower orders are included.

Before symmetry is imposed, the real seed space is
\begin{equation}
  \mathcal W_0(\mathfrak M)
  =
  \operatorname{span}_{\mathbb R}
  \left\{
    Y_\alpha,\ iY_\alpha
  \right\}_{\alpha\in\mathfrak M}.
  \label{app:eq:raw_real_seed_space}
\end{equation}
Both real directions are included because antiunitary operations act antilinearly on state vectors but real-linearly on Hermitian Hamiltonian operators.  Symmetry and Hermitian conjugation may add terms absent from the seed list. For the momentum-independent symmetry matrices and linear momentum maps used here, these operations do not increase the maximum total polynomial degree. The completed operator space determines which terms the model can contain.

\subsection{Symmetry projection}
\label{app:subsec:symmetry_projections}

The continuum symmetry representation acts on a candidate operator as
\begin{equation}
  \mathcal T_g[\mathcal E](\kvec)
  =
  U_g^{\rm cont}
  \left[
    \mathcal E\!\left(\kappa_g(\kvec)\right)
  \right]^{[\eta_g]}
  U_g^{{\rm cont}\dagger}.
  \label{app:eq:operator_action_group_law}
\end{equation}
Although the state representation may carry a projective phase, the phase cancels between the left and right factors in Eq.~\eqref{app:eq:operator_action_group_law}.  Hence \(\mathcal T_g\) is an ordinary group action on Hamiltonian operators.

Let \(\mathcal W_{\rm cl}(\mathfrak M)\) denote the smallest finite real space containing \(\mathcal W_0(\mathfrak M)\) and closed under the selected symmetry operations and Hermitian conjugation. Its Hermitian invariant part is obtained by applying the symmetry average and Hermitian completion directly to the original seeds. Separately, the finite \(Q\) basis must be closed under the selected symmetry maps, but need not be closed under every Fourier transfer.

On this closed space, Hermitian and group projections are
\begin{equation}
  \mathcal P_{\rm H}\mathcal E
  =
  \frac{\mathcal E+\mathcal E^\dagger}{2},
  \qquad
  \mathcal P_{\mathcal G}\mathcal E
  =
  \frac{1}{|\mathcal G|}
  \sum_{g\in\mathcal G}\mathcal T_g[\mathcal E].
  \label{app:eq:term_reynolds_operator}
\end{equation}
The second expression is the Reynolds average.  Because \(\mathcal T_h\mathcal T_g=\mathcal T_{hg}\) on Hamiltonian operators, left multiplication only permutes the elements in the finite sum:
\begin{equation}
  \mathcal T_h\mathcal P_{\mathcal G}\mathcal E
  =\frac{1}{|\mathcal G|}\sum_{g\in\mathcal G}
  \mathcal T_{hg}\mathcal E
  =\mathcal P_{\mathcal G}\mathcal E,
  \qquad
  \mathcal P_{\mathcal G}^2=\mathcal P_{\mathcal G}.
  \label{app:eq:reynolds_projection_proof}
\end{equation}
Thus its image is exactly the fixed space of the group action.  Within the Hermitian closed space, that fixed space is
\begin{equation}
  \mathcal V_{\rm sym}(\mathfrak M)
  =
  \left\{
    \mathcal E\in\mathcal W_{\rm cl}(\mathfrak M):
    \mathcal E^\dagger=\mathcal E,\quad
    \mathcal T_g\mathcal E=\mathcal E
    \ \forall g\in\mathcal G
  \right\}.
  \label{app:eq:symmetry_allowed_operator_space}
\end{equation}

Applying the two projections to a spanning set generally produces zero and linearly dependent matrices. MoireKP determines the independent operators from their sparse polynomial coefficients. A separate sampling matrix contains all their matrix elements at the coefficient-extraction momenta; its rank checks whether those momenta distinguish the independent operators. Symbolically,
\begin{equation}
  \left\{\widetilde Y_j\right\}
  =
  \operatorname{indep}_{\mathbb R}
  \left\{
    \mathcal P_{\rm H}\mathcal P_{\mathcal G}Y_\alpha,\,
    \mathcal P_{\rm H}\mathcal P_{\mathcal G}(iY_\alpha)
  \right\}_{\alpha\in\mathfrak M}.
  \label{app:eq:independent_allowed_operators}
\end{equation}
For the rank check, each operator is normalized by the norm of its polynomial coefficients. Let \(\mathsf A_{\rm n}\) contain the weighted real and imaginary matrix elements of these normalized operators at the coefficient-extraction momenta, and let \(G_{\rm c}=V\Lambda V^T\) be their coefficient Gram matrix. The sampling matrix used for the rank check is \(\mathsf A_{\rm n}V_+\Lambda_+^{-1/2}\), where \(+\) retains eigenvalues above \(\tau_G\). The threshold \(\tau_A\) applies to its singular values, and \(\kappa_A\) is the ratio of the largest to the smallest retained singular value.

For all six models, the sampled matrix rank equals the independently determined operator-space dimension. Table~\ref{tab:final_rank_diagnostics} reports the coefficient counts, ranks, thresholds and condition numbers. For ZrS$_2$ \(M_1\), singular-value decomposition in coefficient space and sampling at additional two-dimensional momenta give the same rank. Appendix~D orthonormalizes these operators under the matrix inner product defined there and extracts their coefficients from \(H_{\rm eff}\).

\section{Coefficient extraction and model selection}
\label{app:model_parameter_determination}

\subsection{Trace projection}
\label{app:subsec:fixed_truncation_projection}

For a fixed truncation \(\mathfrak M\), Appendix~C provides independent Hermitian operators \(\{\widetilde Y_j(\kvec)\}\).  Their nonzero matrix elements define the support \(\Omega_{\mathfrak M}\), and \(\Pi_{\Omega_{\mathfrak M}}\) sets all matrix elements outside that support to zero.  The omitted part of \(H_{\rm eff}\) cannot be changed by determining the coefficients:
\begin{equation}
  \begin{aligned}
  \|H_{\rm cont}-H_{\rm eff}\|_{\rm F}^2
  ={}&
  \|H_{\rm cont}
  -\Pi_{\Omega_{\mathfrak M}}H_{\rm eff}\|_{\rm F}^2\\
  &+
  \|(I-\Pi_{\Omega_{\mathfrak M}})H_{\rm eff}\|_{\rm F}^2 .
  \end{aligned}
  \label{app:eq:omitted_support_decomposition}
\end{equation}
The second term is fixed for a chosen \(\mathfrak M\), although it changes when the truncation is enlarged.

Let \(\mathcal K_{\rm proj}=\{\kvec_i\}_{i=1}^{N_{\rm proj}}\) be the user-specified projection set.  At each point, the columns of \(V_i\) are the \(r_i\) orthonormal eigenstates of \(H_{\rm eff}(\kvec_i)\) in the selected low-energy window.  We write \(\overline A_i=\Pi_{\Omega_{\mathfrak M}}A(\kvec_i)\), and similarly for \(\overline B_i\), and set \(P_i=V_iV_i^\dagger\).  The matrix inner product is
\begin{equation}
  \langle A,B\rangle_{\mathfrak M}
  =\frac{1}{N_{\rm proj}}\sum_i\operatorname{Re}\left\{
  \frac{\operatorname{Tr}(\overline A_i^\dagger\overline B_i)}{N_{\rm cont}^2}
  +\frac{\lambda_1\operatorname{Tr}(P_i\overline A_i^\dagger\overline B_i)}{N_{\rm cont}r_i}
  +\frac{\lambda_2\operatorname{Tr}(P_i\overline A_i^\dagger P_i\overline B_i)}{r_i^2}
  \right\}.
  \label{app:eq:sampled_hilbert_schmidt_inner_product}
\end{equation}
Here \(\lambda_1,\lambda_2\geq0\) weight the support-restricted Hamiltonian action on the target states and its projection within their subspace. Setting both to zero gives the Hilbert--Schmidt trace inner product; nonzero values emphasize the selected low-energy space.

We write \(\|A\|_{\mathfrak M}^2=\langle A,A\rangle_{\mathfrak M}\).  For real coefficients \(\vartheta_j\) in the independent invariant basis, the matrix loss is
\begin{equation}
  \begin{aligned}
  \mathcal L_{\rm mat}(\boldsymbol\vartheta)
  &=\left\|
  \sum_j\vartheta_j\widetilde Y_j-H_{\rm eff}
  \right\|_{\mathfrak M}^2\\
  &=\boldsymbol\vartheta^T G\boldsymbol\vartheta
  -2\boldsymbol\vartheta^T\mathbf b
  +\|H_{\rm eff}\|_{\mathfrak M}^2,\\
  G_{ij}&=\langle\widetilde Y_i,\widetilde Y_j\rangle_{\mathfrak M},\\
  b_i&=\langle\widetilde Y_i,H_{\rm eff}\rangle_{\mathfrak M}.
  \end{aligned}
  \label{app:eq:sampled_projection_loss}
\end{equation}
Stationarity gives \(G\boldsymbol\vartheta=\mathbf b\).  For positive-definite \(G\), a stable factorization orthonormalizes the operators and gives
\begin{equation}
  \begin{gathered}
  Y_a=\sum_j\widetilde Y_j Z_{ja},\qquad
  \langle Y_a,Y_b\rangle_{\mathfrak M}=\delta_{ab},\\
  \theta_a=\langle Y_a,H_{\rm eff}\rangle_{\mathfrak M},\qquad
  H_{\rm cont}^{(\mathfrak M)}(\kvec)=\sum_a\theta_aY_a(\kvec).
  \end{gathered}
  \label{app:eq:trace_projection_coefficients}
\end{equation}
In this orthonormal basis, the Gram matrix is the identity, so the normal equation reduces to the trace coefficient above.  Collecting the real and imaginary parts of the same matrix elements and applying the same weights gives an equivalent linear least-squares problem.  The linear solve and trace extraction are therefore equivalent forms of the same matrix projection.  This applies the trace construction of Ref.~\cite{zhang_universal_2024} simultaneously at several momenta.

\begin{table*}[tbp]
\centering\small
\setlength{\tabcolsep}{7pt}
\caption{Operator-space and sampling-matrix ranks for the six models. $N_{\rm ch}$ is the number of real coefficients, $r_G$ their polynomial span rank, and $r_A$ the sampling-matrix rank. $\tau_G$ and $\tau_A$ are the eigenvalue and singular-value thresholds. $\kappa_A$ is the condition number of $\mathsf A_{\rm n}V_+\Lambda_+^{-1/2}$, defined in Appendix~\ref{app:subsec:symmetry_projections}.}
\label{tab:final_rank_diagnostics}
\begin{tabular}{lrrrrrr}
\toprule
Model & $N_{\rm ch}$ & $r_G$ & $r_A$ & $\tau_G$ & $\tau_A$ & $\kappa_A$ \\
\midrule
MoTe$_2$ $K$ & 156 & 154 & 154 & $4.79\times10^{-10}$ & $4.05\times10^{-8}$ & 219.9 \\
ZrS$_2$ $M_1$ & 479 & 446 & 446 & $6.11\!\times\!10^{-10}$ & $3.75\!\times\!10^{-8}$ & 14.2 \\
ZrS$_2$ $\Gamma$ & 630 & 630 & 630 & $1.98\times10^{-9}$ & $6.95\times10^{-7}$ & 432.3 \\
Trilayer $K_A$ & 188 & 188 & 188 & $2.48\times10^{-10}$ & $1.24\times10^{-9}$ & 1022.8 \\
Trilayer $K_B$ & 34 & 34 & 34 & $6.21\times10^{-11}$ & $4.30\times10^{-10}$ & 976.5 \\
Trilayer $\Gamma$ & 440 & 440 & 440 & $5.34\times10^{-10}$ & $6.55\times10^{-9}$ & 39.6 \\
\bottomrule
\end{tabular}
\end{table*}

\subsection{Nonlinear refinement}
\label{app:subsec:nonlinear_refinement}

All six models use linear matrix-coefficient extraction. The optional nonlinear band-energy refinement described below was not used. The solution \(\boldsymbol\vartheta^{(0)}\) of Appendix~D.1 can initialize such a refinement without changing the operators, harmonic shells, or momentum-expansion orders. Let \(\mathcal K_{\rm band}\) denote the chosen momenta and \(\mathcal I_i\) the target bands at each point. The refined coefficients minimize
\begin{equation}
  \mathcal L_{\rm nl}(\boldsymbol\vartheta)
  =\mathcal L_{\rm mat}(\boldsymbol\vartheta)
  +\frac{\lambda_b}{N_{\rm band}}
  \sum_{i\in\mathcal K_{\rm band}}
  \frac{1}{|\mathcal I_i|}
  \sum_{n\in\mathcal I_i}
  \left[
    \varepsilon_{in}^{\rm cont}(\boldsymbol\vartheta)
    -\varepsilon_{in}^{\rm eff}
  \right]^2.
  \label{app:eq:nonlinear_refinement_loss}
\end{equation}
Here \(N_{\rm band}=|\mathcal K_{\rm band}|\) is the number of sampled momenta and \(\lambda_b\geq0\) sets the relative band weight.

For a nondegenerate band, the analytic derivative is given by the Hellmann--Feynman relation~\cite{feynman_Forces_1939},
\begin{equation}
  \frac{\partial\varepsilon_{in}^{\rm cont}}
       {\partial\vartheta_j}
  =
  \left\langle
    u_{in}^{\rm cont}
  \middle|
    \widetilde Y_j(\kvec_i)
  \middle|
    u_{in}^{\rm cont}
  \right\rangle .
  \label{app:eq:hellmann_feynman_coefficient_jacobian}
\end{equation}
The optional nonlinear solver uses these derivatives only at samples with nondegenerate target bands. Before refinement, it checks the reference and initial model spectra and excludes a momentum from the band-energy loss if a gap touching a selected band, including a gap across the selected-window boundary, is at most \(10^{-8}\) eV by default. The retained band-sample set is then fixed, while the matrix-fit samples remain unchanged. If no band samples remain, an energy-alignment anchor is degenerate, or a retained sample develops a degeneracy during refinement, the refinement is skipped or stopped and the input coefficients are retained.

\subsection{Automatic selection of harmonic shells}
\label{app:subsec:truncation_selection}

The automatic scan selects the numbers of intralayer and interlayer harmonic shells at fixed user-specified momentum-expansion orders. It is omitted when the shell numbers are supplied.

When the shells are not specified, MoireKP forms candidate pairs \(c=(n_{\rm intra},n_{\rm inter})\) from the distinct momentum-transfer shells present in the selected plane-wave basis.  The search may instead be restricted by a user-supplied list of candidate pairs or a maximum shell. For each candidate, a Hermitian mask \(M_c\) selects the local terms and the requested intralayer and interlayer shells.  The Hamiltonian used in the scan is
\begin{equation}
  H_{\rm eff}^{(c)}(\kvec)
  =
  M_c\odot H_{\rm eff}(\kvec),
  \label{app:eq:harmonic_ablation}
\end{equation}
where \(\odot\) denotes elementwise multiplication.  The scan uses an evenly spaced subset of the momenta for which \(H_{\rm eff}\) has been constructed.

At each selected momentum, the program diagonalizes \(H_{\rm eff}^{(c)}\) and the unmasked \(H_{\rm eff}\).  It compares the target-band energies and the corresponding low-energy subspaces.  For an \(r\)-dimensional target space, the subspace overlap is
\begin{equation}
  F_c(\kvec)
  =
  \frac{1}{r}
  \operatorname{Tr}
  \left[\Pi_c(\kvec)\Pi_{\rm eff}(\kvec)\right],
  \label{app:eq:harmonic_subspace_overlap}
\end{equation}
Here \(\Pi_c\) and \(\Pi_{\rm eff}\) project onto the target-band subspaces of the masked and unmasked effective Hamiltonians, respectively. Both act in the effective-model coefficient space, whereas \(P\) in Eq.~\eqref{eq:embedded_coordinate_anchors} selects the low-energy space in TAPW coordinates. The projected difference \(\Pi_{\rm eff}[H_{\rm eff}^{(c)}-H_{\rm eff}]\Pi_{\rm eff}\) compares Hamiltonian matrix elements within the target subspace.

By default, the scan requires a band root-mean-square difference no larger than \(1\,\mathrm{meV}\) and a mean subspace overlap above \(0.90\), and chooses the candidate with the fewest shells.  MoireKP then constructs the allowed operators and determines their coefficients from Eq.~\eqref{app:eq:trace_projection_coefficients} on \(\mathcal K_{\rm proj}\).

The displayed band paths share momenta with harmonic selection and coefficient extraction; no separate path is withheld for validation.

\subsection{Model parameters and numerical comparisons}
\label{app:subsec:paper_model_records}

Table~\ref{tab:paper_model_configurations} collects the construction parameters and target-window path comparisons. The coefficient-extraction points are specified in the model configurations and are also used for the rank checks in Appendix~\ref{app:subsec:symmetry_projections}. The reference basis is defined by Eq.~\eqref{eq:embedded_coordinate_anchors}; energy errors and embedded-state overlaps follow Sec.~\ref{sec:validation_protocol}. No nonlinear refinement was used for any of the six models.

\moirekp{} can automatically determine the number of low-energy states by enumerating complete multiplets from the target edge within a specified candidate sequence. Selection uses the sector-dependent criteria below. Accepted non-$\Gamma$ candidates are ranked first by smaller dimension, then by smaller band RMS, larger reference overlap, smaller symmetry residual when available, and finally by candidate label. The $\Gamma$ search expands its representative cluster until the local criteria are met. Reference-frame covariance is checked separately after selection and is not a search acceptance criterion. The search raises an error if no candidate satisfies the required criteria. Alternatively, the selected states and reference columns can be specified by the user.

The non-$\Gamma$ branches compare bands at both $\Gamma$ endpoints and the midpoint of $M$--$K$ and require anchor $\sigma_{\min}\geq0.05$. Band RMS/maximum differences are limited to 10/30 meV for bilayer $K$ and $M_1$; they are reported but are not used to reject candidates for the partial trilayer $K_A/K_B$ sectors. The two $\Gamma$ branches compare eight bands at $\Gamma$, requiring RMS and maximum differences below 3 meV and anchor $\sigma_{\min}^2\geq0.05$. These are selection thresholds, not the final errors or overlaps in Table~\ref{tab:paper_model_configurations}. Multiplets use a $10^{-5}$ eV tolerance; non-$\Gamma$ candidates use a 0.1 eV active-layer window and a maximum local dimension of 16. The $\Gamma$ target-window degeneracy tolerance is 0.01 meV. References are chosen at $\Gamma$ and reused across channels, each of which is checked for numerical rank.

For wave-function overlaps, bilayer MoTe$_2$ \(K\) uses energy-ordered pairs; the other models use one-to-one maximum-overlap matching within fixed TAPW target sets. These are edge-band sets for ZrS$_2$ \(M_1/\Gamma\) and trilayer \(\Gamma\). The trilayer \(K_A/K_B\) sets are fixed by a joint one-to-one assignment to full-spin TAPW states across the crossing.

The bilayer MoTe$_2$ calculation uses a 1302-atom supercell. Real-space H/S averaging uses its relaxed coordinates and six spatial operations identified at a 0.05~\AA\ tolerance; TAPW uses the corresponding rigid reference structure. The subsequent TAPW symmetry search uses a 0.005~\AA\ site tolerance and retains \(C_{3z}\) and \(C_2\mathcal T\).

For trilayer MoTe$_2$, the spin-block approximation is tested before low-energy reduction by comparing the full-spin TAPW spectrum with the spin-diagonal-block spectra. A joint one-to-one state assignment is used, and each spectrum is aligned at its \(K_A\) valence-band maximum, with the same shift applied to \(K_B\). The resulting energy changes are reported in Sec.~\ref{sec:aab_mote2_multivalley}.

\begin{table*}[tbp]
\centering
\small
\caption{Construction parameters and path validation of the six continuum models in Figs.~\ref{fig:mote2_k_flagship}--\ref{fig:mote2_aab_multivalley}. $N_{\rm local}$ counts the internal reference states and $N_{\rm cont}$ is the final matrix dimension [Eq.~\eqref{eq:continuum_composite_index}]. Energy differences are in meV after the alignment specified in Sec.~\ref{sec:validation_protocol}, using the same target sets. Reference energies are rounded for display. Overlaps use the embedding and cluster definition in Sec.~\ref{sec:validation_protocol}.}
\label{tab:paper_model_configurations}
\setlength{\tabcolsep}{2.5pt}
\renewcommand{\arraystretch}{1.15}
\begin{tabular*}{\textwidth}{@{\extracolsep{\fill}}lcccccc}
\toprule
 & \multicolumn{1}{c}{Bilayer MoTe$_2$} & \multicolumn{2}{c}{Bilayer ZrS$_2$} & \multicolumn{3}{c}{Trilayer MoTe$_2$} \\
Model & $K$ & $M_1$ & $\Gamma$ & $K_A$ & $K_B$ & $\Gamma$ \\
\midrule
\multicolumn{7}{l}{\textit{Construction parameters}} \\
$N_{\rm local}/N_{\rm cont}$ & 2/54 & 4/88 & 8/152 & 2/36 & 1/18 & 4/76 \\
$N_Q$ per sector & 27,27 & 22,22 & 19,19 & 18,18 & 18 & 19,19 \\
Reference states & \shortstack{$d_+\uparrow$\\layers 1,2} & \shortstack{layer Kramers\\pairs} & $a_{\ell sm}^{(+)}$ & \shortstack{$d_+\uparrow$\\$A_1,A_2$} & \shortstack{$d_-\downarrow$\\$B$} & \shortstack{$d_{z^2}\uparrow\!\downarrow$\\$A_1,A_2$} \\
TAPW cutoff $n_g$ & 6 & 7 & 4 & 4 & 4 & 4 \\
Harmonic shells (diagonal, off-diagonal) & 1,3 & 1,2 & 1,1 & 1,2 & 1,0 & 3,1 \\
Orders $(m_{\rm kin},m_{\rm intra},m_{\rm inter})$ & 10,5,5 & 6,4,6 & 8,4,4 & 10,4,5 & 10,2,0 & 8,2,6 \\
$E_{\rm ref}$ (eV) & $-4.61$ & $-5.22$ & $-6.21$ & $-4.63$ & $-4.68$ & $-4.72$ \\
\midrule
\multicolumn{7}{l}{\textit{Validation results}} \\
Target bands & top 8 & bottom 8 & top 10 & top 6 & top 3 & top 8 \\
TAPW $\to H_{\rm eff}$ RMS (meV) & $<0.01$ & 0.02 & 0.05 & 1.12 & 1.55 & $<0.01$ \\
$H_{\rm eff}\to$ continuum RMS (meV) & 0.82 & 0.64 & 0.62 & 0.75 & 0.76 & 0.64 \\
TAPW $\to$ continuum RMS (meV) & 0.82 & 0.64 & 0.62 & 0.98 & 1.07 & 0.65 \\
TAPW $\to$ continuum maximum (meV) & 1.98 & 1.37 & 2.74 & 3.00 & 2.47 & 2.56 \\
Mean / minimum $\mathcal O$ & 0.98 / 0.81 & 0.92 / 0.90 & 0.99 / 0.97 & 0.90 / 0.50 & 0.83 / 0.49 & 0.99 / 0.95 \\
\bottomrule
\end{tabular*}
\par\smallskip
\begin{minipage}{\textwidth}
\small\raggedright
Construction and model comparison use 61 points on $\Gamma-M-K-\Gamma$. Statistics give equal weight to the samples in each target window. $K_A/K_B$ use joint one-to-one TAPW target assignments.

Harmonic counts are distinct nonzero transfer-radius shells in the final symmetry/adjoint-closed support; zero transfer is excluded and radii are grouped to $10^{-8}$~\AA$^{-1}$. The two diagonal blocks have equal counts where both exist. Reference states are layer-resolved for non-$\Gamma$ models and joint same-$Q$ states for $\Gamma$ models; $a_{\ell sm}^{(+)}$ is the symmetric S $p_x/p_y$ spinful reference of Eq.~\eqref{app:eq:zrs2_reference_orbitals}.

Low-energy dimensions, reference states and harmonic truncations were selected automatically; cutoffs, orders, $E_{\rm ref}$ and coefficient-extraction points were user inputs. Coefficient extraction for ZrS$_2$ $\Gamma$ uses $(\lambda_1,\lambda_2)=(2.25,22.5)$ on the highest 20 states; the other five models use unweighted extraction [Eq.~\eqref{app:eq:sampled_hilbert_schmidt_inner_product}].
\end{minipage}
\end{table*}

For the normalized geometric fields \(f\) shown in Figs.~\ref{fig:mote2_k_flagship}--\ref{fig:mote2_aab_multivalley}, the mean relative differences \(\langle|f_{\rm cont}-f_{\rm TAPW}|\rangle/\langle|f_{\rm TAPW}|\rangle\), averaged uniformly over the first moir\'e Brillouin zone, are 2.7\% and 3.4\% for bilayer MoTe$_2$ \(K\) Berry curvature and quantum metric, 2.7\% for the ZrS$_2$ \(\Gamma\) quantum metric, and 6.6\% and 7.4\% for trilayer \(K_A\) Berry curvature and \(\Gamma\) quantum metric, respectively.

\end{document}